\documentclass[twocolumn]{aastex631}
\usepackage{ulem}
\usepackage{mathtools}

\graphicspath{ {./} }

\begin{document}

\title{Formation of Stripped Stars From Stellar Collisions in Galactic Nuclei} 

\author[0009-0003-8690-8297]{Charles F.\ A.\ Gibson}
\affiliation{Department of Physics, Allegheny College, Meadville, Pennsylvania 16335, USA}

\author[0000-0003-4412-2176]{Fulya K{\i}ro\u{g}lu}
\affiliation{Center for Interdisciplinary Exploration \& Research in Astrophysics (CIERA) and Department of Physics \& Astronomy \\ Northwestern University, Evanston, IL 60208, USA}

\author[0000-0002-7444-7599]{James C.\ Lombardi Jr.}
\affiliation{Department of Physics, Allegheny College, Meadville, Pennsylvania 16335, USA}

\author[0000-0003-0984-4456]{Sanaea C.\ Rose}
\affiliation{Center for Interdisciplinary Exploration \& Research in Astrophysics (CIERA) and Department of Physics \& Astronomy \\ Northwestern University, Evanston, IL 60208, USA}

\author[0009-0009-6575-2207]{Hans D.\ Vanderzyden}
\affiliation{Department of Physics, Allegheny College, Meadville, Pennsylvania 16335, USA}

\author[0000-0001-6350-8168]{Brenna Mockler}
\affiliation{The Observatories of the Carnegie Institution for Science, Pasadena, CA 91101, USA}

\author[0000-0003-0648-2402]{Monica Gallegos-Garcia}
\affiliation{Center for Interdisciplinary Exploration \& Research in Astrophysics (CIERA) and Department of Physics \& Astronomy \\ Northwestern University, Evanston, IL 60208, USA}

\author[0000-0002-4086-3180]{Kyle Kremer}
\affiliation{TAPIR, California Institute of Technology, Pasadena, CA 91125, USA}

\author[0000-0003-2558-3102]{Enrico Ramirez-Ruiz}
\affiliation{Department of Astronomy and Astrophysics, UCO/Lick Observatory, University of California, 1156 High Street, Santa Cruz, CA 95064, USA}

\author[0000-0002-7132-418X]{Frederic A.\ Rasio}
\affiliation{Center for Interdisciplinary Exploration \& Research in Astrophysics (CIERA) and Department of Physics \& Astronomy \\ Northwestern University, Evanston, IL 60208, USA}

\begin{abstract}
Tidal disruption events (TDEs) are an important way to probe the properties of stellar populations surrounding supermassive black holes. Observed spectra of several TDEs, such as ASASSN-14li, show high nitrogen to carbon abundance ratios, leading to questions about their progenitors. 
Disrupting an intermediate- or high-mass star that has undergone CNO processing, increasing the nitrogen in its core, could lead to an enhanced nitrogen TDE.
Galactic nuclei present a conducive environment for high-velocity stellar collisions that can lead to high mass loss, stripping the carbon- and hydrogen-rich envelopes of the stars and leaving behind the enhanced nitrogen cores.
TDEs of these stripped stars may therefore exhibit even more extreme nitrogen enhancement.
Using the smoothed particle hydrodynamics code {\tt StarSmasher}, we provide a parameter space study of high-velocity stellar collisions involving intermediate-mass stars, analyzing the composition of the collision products. We conclude that high-velocity stellar collisions can form products that have abundance ratios similar to those observed in the motivating TDEs. Furthermore, we show that stars that have not experienced high CNO processing can yield low-mass collision products that retain even higher nitrogen to carbon abundance ratios.
We analytically estimate the mass fallback for a typical TDE of several collision products to demonstrate consistency between our models and TDE observations.
Lastly, we discuss how the extended collision products, with high central to average density ratios, can be related to repeated partial TDEs like ASASSN-14ko and G objects in the Galactic Center.
\end{abstract}

\section{Introduction} \label{sec:intro}

Most galactic nuclei harbor supermassive black holes (SMBHs), surrounded by dense stellar clusters \citep[e.g.,][]{Ghez+03,Genzel+03,FerrareseFord05,Schodel+18}. Interactions within the cluster can drive stars onto nearly radial orbits about the SMBH such that the stars pass within their tidal limit and are destroyed \citep[e.g.,][]{Hills1975,Rees1988,Alexander99,MagorrianTremaine99,WangMerritt04}. These tidal disruption events (TDEs) produce electromagnetic signatures as stellar material is later accreted onto the SMBH \citep[e.g.,][]{GuillochonRR13,Dai+21}, and spectra from TDEs offer a powerful probe of the stellar populations in galactic nuclei \citep[e.g.,][]{Yang17,Mockler+22,Miller+23}.

Observed spectra of several recent TDEs, including ASASSN-14li, PTF15af, and iPTF16fnl, display intriguingly large nitrogen-to-carbon abundance ratios \citep{Cenko+16,Kochenek+16,Blagorodnova+17,Blagorodonva+19,Yang17}.
Observational lower limits range from $10^{1.5}$ to $10^{2.4}$ times the corresponding solar value ([N/C]= 1.5 to 2.4) \citep{Yang17,Miller+23}.
Massive stars ($m\gtrsim3\,M_{\odot}$)
have been proposed as a potential progenitor for these TDEs \citep{Mockler+22,Miller+23}.
However, in the case of ASASSN-14li, the constraints on [N/C] suggest that a more plausible progenitor is a star that has experienced significant CNO burning and has been subsequently stripped of its hydrogen-rich envelope \citep{Miller+23}.
Observations of repeated partial TDEs, such as ASASSN-14ko, may also be explained by a star with an extended, low-mass envelope and a high core-to-envelope mass ratio \citep{Liu+23,2024arXiv240601670L}, expected from stripping of an evolved giant \citep{2013ApJ...777..133M} or from a high-velocity collision product. A star like this might have either formed close to the SMBH or been placed in a tight orbit around the SMBH following a stellar collision.

One natural way to produce stripped stars is Roche lobe overflow in binary systems \citep[and references therein]{Goettberg+18}.
However, other channels can often mimic the same evolutionary effects as binary interactions.
For example, stars can be stripped of their envelopes by ram pressure as they pass through the disks of active galactic nuclei \citep{McKernan+22}.
Additionally, high-velocity stellar collisions that produce significant mass loss may leave stars that mimic the composition and structure of stripped stars formed through binary evolution \citep[e.g.][]{LaiRasioShapiro93, Freitag+08,Rauch99}.
Regardless of their specific formation mechanisms, stars stripped of their envelopes will have outer layers that contain elevated helium and nitrogen abundances \citep{Goettberg+18,Mockler+24}.
These stripped stars have been invoked to explain hydrogen-poor supernovae \citep{WheelerLevreault85,Podsiadlowski+92} and TDEs with elevated nitrogen levels \citep{Gallegos-Garcia+18, Mockler+22,Miller+23,Mockler+24}.

Here we focus on the stellar collision scenario for forming stripped stars. Galactic nuclei are natural locations for both stellar collisions and subsequent TDEs, given their high stellar densities in the proximity of a central SMBH.
The extreme stellar density and high velocity dispersion in galactic nuclei may give rise to the formation of unusual collision products such as G~objects and exotic massive stars following nonstandard evolution pathways \citep{Rose+23}.
Moreover, the tidal breakup of binary stars in galactic nuclei such as our own Galactic Center can eject one companion as a hypervelocity star and leave the other in a tight orbit around the SMBH \citep{2011ApJ...731..128A}.
These captured stars form a population that, in steady state, can undergo collisions at a rate comparable to their capture rate; such highly-energetic collisions can potentially strip and even destroy the stars, producing supernova-like events \citep{Balberg+13}.
\citet{LaiRasioShapiro93} and  \citet{Freitag+08} modeled these high-velocity stellar collisions between main sequence (MS) stars with smoothed particle hydrodynamics, focusing on the mass loss from the parent stars, and \citet{Ryu24} predicted the observational signatures of collisions between red giants through modeling the interactions with {\tt AREPO} \citep{AREPO16,AREPO20} and {\tt MESA} \citep{Paxton+18}.
In this work, we aim to provide a compositional analysis of the products from high-velocity stellar collisions, so we can compare their chemical composition to that of the observed high [N/C] TDEs.

Our paper is organized as follows.
In Section~\ref{sec:rates} we provide an introduction to the rates of stellar collisions in galactic nuclei.
Section~\ref{sec:methods} describes our numerical methodology, while Section~\ref{sec:results} presents our hydrodynamic simulation results.
In Section \ref{sec:implications} we discuss several implications of the post-collision stripped stars.
We determine the chemical composition of the fallback material from a total TDE of several post-collision stripped stars and discuss properties of extended collision products in relation to observations of repeated partial tidal disruption events and of G objects in the Galactic Center.
In the Appendix, we give more details about how our parent star models are generated and we provide additional information on the stellar profiles of the collision products.

\section{Rates of Collisions in Galactic Nuclei} \label{sec:rates}

\begin{figure*}
    \centering
    \includegraphics[width=0.497\linewidth]{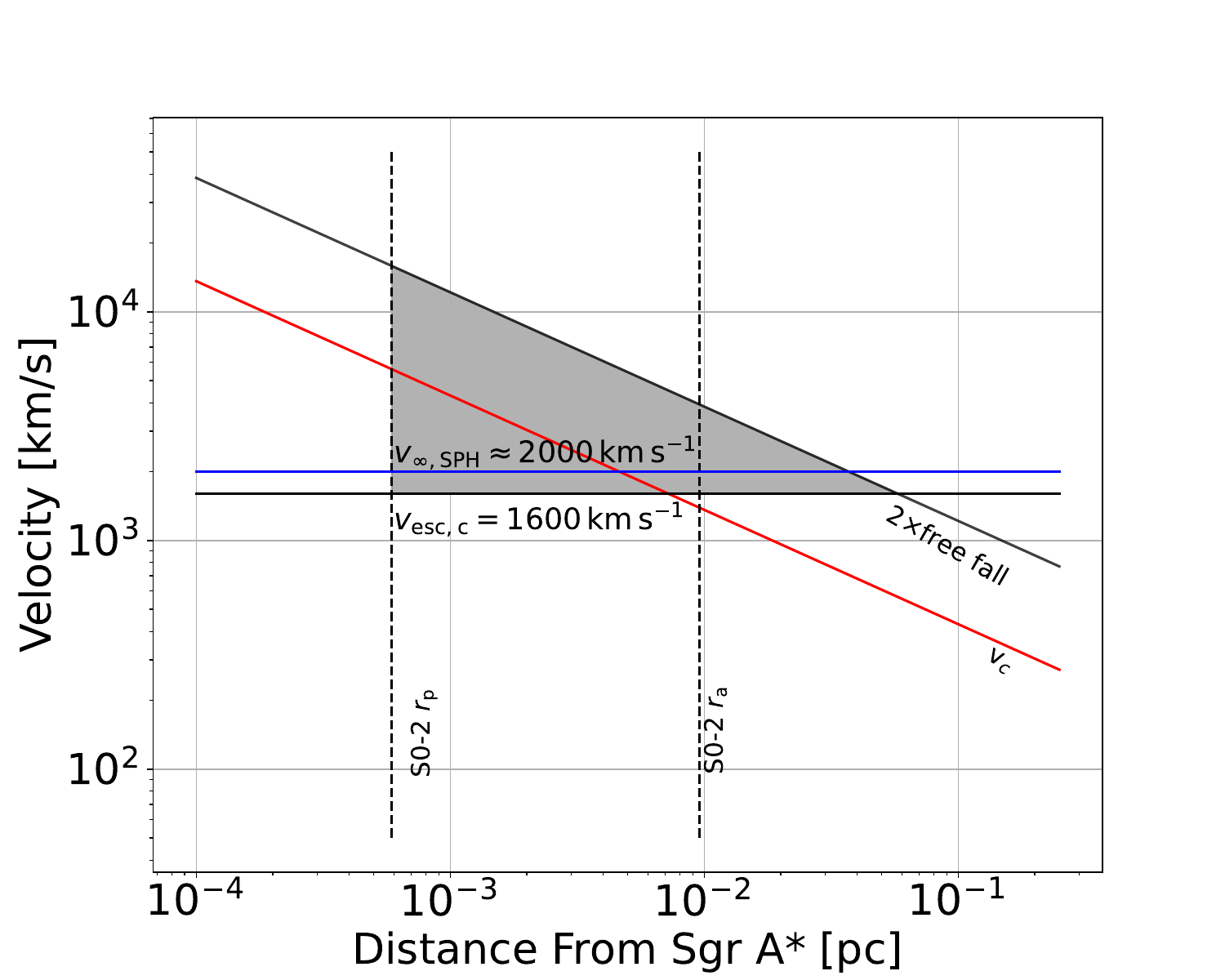}
    \includegraphics[width=0.497\linewidth]{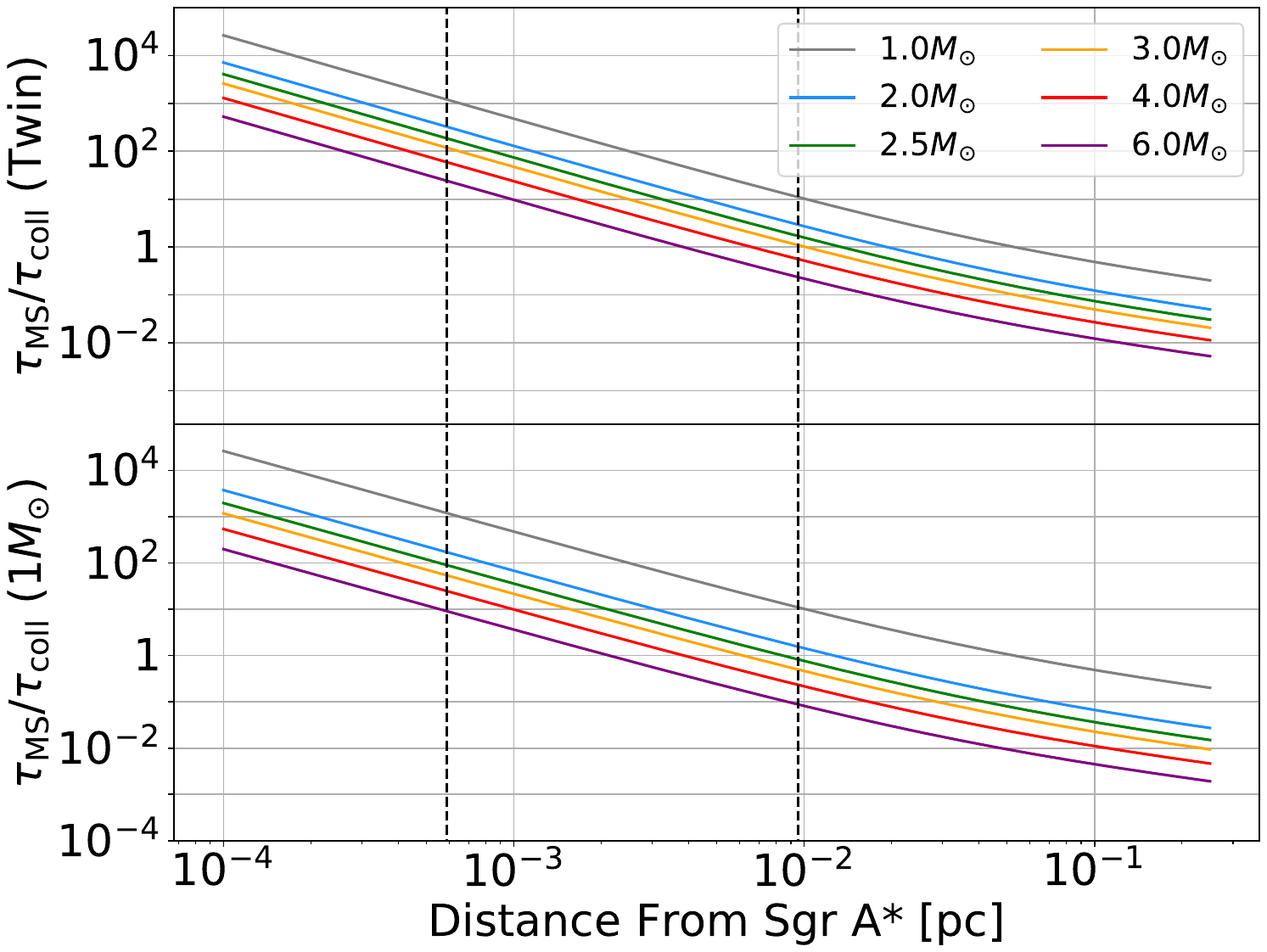}
    \caption{\textbf{(Left)} The possible velocity at infinity values and collision rates within the inner parsec of the Milky Way. The left panel shows the orbital velocities of stars at a distance $R$ from the center of the Milky Way. The red line is the Keplerian circular velocity $v_{\rm c}$ of a star at a given distance. The solid black line is a proposed upper limit of the relative velocity at infinity: this velocity, given by $v=2\sqrt{2GM_{\bullet}/R}$, describes two stars on parabolic orbits that collide at their periapses. The horizontal black line at 1600\,km\,s$^{-1}$ denotes the escape velocity of a point mass at the center of a typical MS star (see Footnote~\ref{fn:vesc_c}). This value provides an estimate for the velocity required to strip substantial mass from the MS sequence stars we consider. The horizontal blue line at 2000\,km\,s$^{-1}$ roughly denotes the lowest velocity required to strip a MS star as predicted from SPH simulations from our models and those of \citet{FreitagBenz2005}. The periapsis and apoapsis of S0-2, the closest known star to Sgr A*, are shown for reference. Within this constraint, both the circular velocity and upper limit of relative $v_{\infty}$ are greater than the estimated stripping velocity. \textbf{(Right)} The ratio of the MS lifetime of the star of interest to the collision timescale of that star.
    The top panel assumes that the background stars are all identical to the star of interest, namely having a mass of 1, 2, 2.5, 3, 4, and 6 $M_{\odot}$ for each curve respectively. If this is the case, it is reasonable to expect at least one collision between two stars of mass $\lesssim3\,M_{\odot}$ inside $\sim 10^{-2}$ pc. The bottom right plot assumes a constant background star of $1M_{\odot}$. With these constraints, we expect that a star of mass $\lesssim2.5M_{\odot}$ will experience a collision during its lifetime within this region of the galaxy. These two panels demonstrate that high-velocity stellar collisions are both possible and likely within the Galactic Center.}
    \label{fig:tcoll}
\end{figure*}

The formation of stripped stars via stellar collisions requires two basic criteria: First, the collisions must be sufficiently energetic to strip the stellar envelopes. Second, the collision timescale must be sufficiently short for collisions to be likely. In this section, we use the Galactic Center to explore the parameter space where these two criteria are met.

In this work we consider stars with masses between $1M_{\odot}$ and $6M_{\odot}$.
We provide an initial estimate of the velocity required to strip all the mass from a MS star to be $v_{\rm rel}\approx1600$ km\,s$^{-1}$, the escape speed from the center of a ``typical'' MS star
in this mass range.\footnote{For the stars considered in this work, the central escape speed monotonically increases with increasing mass from 1371 km\,s$^{-1}$ to 1806 km\,s$^{-1}$ between the $1M_{\odot}$ and $6M_{\odot}$ stars.\label{fn:vesc_c}}
Velocities of this order of magnitude have been shown to strip substantial mass from the stars involved \citep{LaiRasioShapiro93,FreitagBenz2005,Rauch99}, where \cite{FreitagBenz2005} showed that $v_{\rm rel}\gtrsim2000$ km\,s$^{-1}$ can strip $\sim50$\% of the mass of MS stars in a collision.
These extreme velocities are likely to be found in galactic nuclei where the high orbital velocities and velocity dispersion can be well approximated by the Keplerian circular velocity \citep{Alexander99,AlexanderPfuhl14}.
Although the observed TDEs motivating our project occur in other galaxies, we consider our own Galactic Center because it has been widely studied and because it is a fitting environment: indeed, Sagittarius A* (Sgr A*) has just about the right mass to make TDEs likely for MS stars \citep{Stone+16}.
Much less massive black holes (such as intermediate-mass black holes, with $M_{\bullet}\lesssim10^5\,M_{\odot}$) disrupt MS stars too far from their horizons, while much more massive SMBHs ($M_{\bullet}\gtrsim10^8\,M_{\odot}$) swallow MS stars whole, without any TDE signatures \citep{Beloborodov+92,MagorrianTremaine99,2012ApJ...757..134M}.
Indeed the majority of observed TDEs are around SMBHs with masses in the relatively narrow range $10^6 - 10^7\,M_{\odot}$ \citep[e.g.,][]{2019ApJ...872..151M,wevers_black_2019,Hammerstein_integral_2023}.

The left panel of Figure \ref{fig:tcoll} illustrates that velocities comparable to what we will consider here, as in \cite{FreitagBenz2005}, occur within the inner $\sim0.05$\,pc of Sgr A*, with mass $4.3\times10^6\,M_{\odot}$ \citep{GravCollab_Abuter22}.
For comparison, S0-2 (S2), the closest observed star to Sgr A*, has a pericenter distance of $5.8\times10^{-4}\,$pc and an apocenter at $9.6\times10^{-3}\,$pc \citep[e.g.,][]{Gillessen+17}, a region where high enough velocity collisions may occur to strip substantial mass from the stars involved.

The collision timescales within the inner $10^{-2}$\,pc in our Galactic Center are smaller than, or comparable to, the MS lifetime, $\tau_{\rm{MS}}$, of the stars being considered (see Figure~\ref{fig:tcoll}).
The density within the inner parsec of the Milky Way is taken as $\rho(r)=1.35\times10^6(r/0.25{\rm{ pc}})^{-\gamma}\,M_{\odot}\,\rm{pc}^{-3}$ \citep{Genzel+10}, where $\gamma$ is the slope of the density profile, and the number density is given by $n=\rho/m$ where $m$ is the average mass of the background stars in the inner parsec \citep{Rose+23,Rose+24}.
Additionally, the velocity dispersion is given by $\sigma=\sqrt{GM_{\bullet}/(r(1+\gamma))}$ \citep[e.g.,][]{Alexander99,AlexanderPfuhl14,Rose+23,Rose+24}.
These values can be used to calculate the collision timescale
\begin{equation}
        t_{\rm coll}=\left[4\sqrt{\pi}n\sigma r_{\rm coll}\left(1+\frac{G(m_1+m_2)}{\sigma^2r_{\rm coll}}\right)\right]^{-1} \label{equation:tcoll}
\end{equation}
where $r_{\rm coll}=r_1+r_2$ \citep[ignoring tidal effects,][]{BinneyTremaine87}.
We choose $\gamma=1.25$, representative of a much larger range of accepted values from $0.9\lesssim\gamma\lesssim1.75$ \citep{Schodel+18,Panamarev+19,LinialSari22,Gallego-Cano+18,Rose+23}.

The total, integrated collision timescale over the full MS lifetime $\tau_{\rm MS}$ of a star is
\begin{equation}
    \tau_{\rm coll}=\tau_{\rm{MS}}\left(\int_0^{\tau_{\rm MS}}\frac{dt}{t_{\rm{coll}}}\right)^{-1}
\end{equation}
\citep{Dale+09}, where $t=0$ corresponds to the zero-age MS.
This integral accounts for a changing radius (and changing $r_{\rm{coll}}$), and thus a changing $t_{\rm{coll}}$ over the star's lifetime.
For $\tau_{\rm MS}/ \tau_{\rm coll} >1$, the ratio approximates the expected number of collisions that the star will experience during its lifetime. For $\tau_{\rm MS} / \tau_{\rm coll} <1$, the ratio approximates the probability that the star will experience a single collision during its lifetime \citep{Rose+23}.

To estimate collision timescales, we consider two possible scenarios.
First, we assume the stars collide with other identical intermediate-mass stars: specifically, $r_1 = r_2$ and $m_1 = m_2$.
The resulting collision timescales as a function of distance from Sgr A* in this ``twin" scenario are shown for several star masses in the top of the right panel in Figure \ref{fig:tcoll}.
In this case, both stars involved in the collision may yield a stripped star with an elevated N/C abundance ratio due to previous CNO processing.
Second, we assume instead that stars collide with a $1M_{\odot}$ background star, as in \citet{Rose+23}. The resulting collision timescales in this scenario are shown in the bottom of the right panel in Figure \ref{fig:tcoll}.
A $1\,M_\odot$ background star mass is reasonable when considering evidence for a top-heavy IMF in the Galactic Center \citep{Paumard+06,Bartko+10,AlexanderPfuhl14,Lu+09,Lu+13,Hosek+19}
and fits within the background star mass prediction of $m<1.5M_{\odot}$ set in \cite{Gallego-Cano+19}.
Collisions in this second scenario may only be able to yield a single stripped star with an elevated N/C abundance ratio because the $1M_{\odot}$ star will not have experienced significant CNO processing.

The right panel in Figure \ref{fig:tcoll} shows that, regardless of the choice of $m$, the stars considered in this work are likely to experience a collision in the Galactic Center.
For both of choices of background star, a star with a mass $m\lesssim3\,M_{\odot}$ within the orbital constraints of S0-2 is likely to experience at least one collision during its lifetime.

We find that the collision speeds required to strip stars occur in the same locations where we expect frequent stellar collisions.
As shown in Figure \ref{fig:tcoll}, we find that galactic nuclei are an ideal location for the formation of stripped stars via high velocity stellar collisions.
However, it is important to note that other galaxies may have different conditions that may influence the rates and outcomes of these collisions.
\cite{FreitagBenz2005} examine how collisional rates in nuclear star clusters depend on the SMBH mass; they calculate the collision timescales $T_{\rm coll}=(n\sigma v_{\rm rel})^{-1}$, where $n$ is the stellar density, $\sigma$ is the collisional cross section, and $v_{\rm rel}$ is the relative velocity.
For a given distance from the SMBH (which sets $v_{\rm rel}$), they find that the maximum collision timescale occurs for a SMBH mass of $2\times10^6\,M_{\odot}$, not much lower from that of Sgr A*.
For decreasing SMBH masses below $2\times10^6\,M_{\odot}$, gravitational focusing increases the cross-section $\sigma$ quickly enough for $\sigma v_{\rm rel}$ to increase as well.
For increasing SMBH masses above $2\times10^6M_{\odot}$, gravitational focusing plays a lesser role and $\sigma v_{\rm rel}$ increases due to the increasing $v_{\rm rel}$.
Therefore, the collisional timescale is likely even smaller in most other galaxies than it is in the Milky Way.
Although the collision timescale for lower mass galaxies may be lower than that of the Milky Way, the orbital velocities of stars are also lower, meaning that these galaxies might not meet both criteria set forth to create stripped stars dynamically.
However, higher mass galaxies have both higher orbital velocities and higher collision rates, theoretically making them even better candidates for these high velocity collisions.
Furthermore, there is a high proportion of observed TDEs stemming from galaxies with high central light concentrations \citep{Law-Smith+17,Dodd+21,Hammerstein+21,Polkas+23,Dodd+23}.
These high light concentrations, corresponding to high central stellar densities, mean that galaxies with high collision rates may also have high TDE rates of the stellar collision products.
These factors make a compelling argument that many galactic nuclei would allow for the high-velocity stellar collisions being considered in this work.

\section{Methods} \label{sec:methods}

\subsection{Initial {\tt MESA} Evolution} \label{subsec:init_mesa_evo}

In this work, we employ realistic star models created with {\tt MESA} (v22.05.1) \citep{Paxton+11,Paxton+13,Paxton+15,Paxton+18,Paxton+19,Jermyn+23}.\footnote{Our {\tt MESA} models are completely consistent with ones generated by v24.03.1.} Each star has an initial solar metallicity, $Z=0.02$, and an initial helium mass fraction of $Y=0.28$. We track the abundances of eight elements throughout both the {\tt MESA} evolution and {\tt StarSmasher} collision: $^1$H, $^3$He, $^4$He, $^{12}$C, $^{14}$N, $^{16}$O, $^{20}$Ne, and $^{24}$Mg. These quantities are later presented with respect to the solar abundances, where $^1{\rm H}_{\odot}=0.7438$, $^4{\rm He}_{\odot}=0.244$, $^{12}{\rm C}_{\odot}=0.00257$, $^{14}{\rm N}_{\odot}=0.000704$, $^{16}{\rm O}_{\odot}=0.00583$, $^{20}{\rm Ne}_{\odot}=0.00171$, $^{24}{\rm Mg}_{\odot}=0.000633$ \citep{AsplundAmarsiGrevesse20}.

\begin{figure}
        \includegraphics[width=\linewidth]{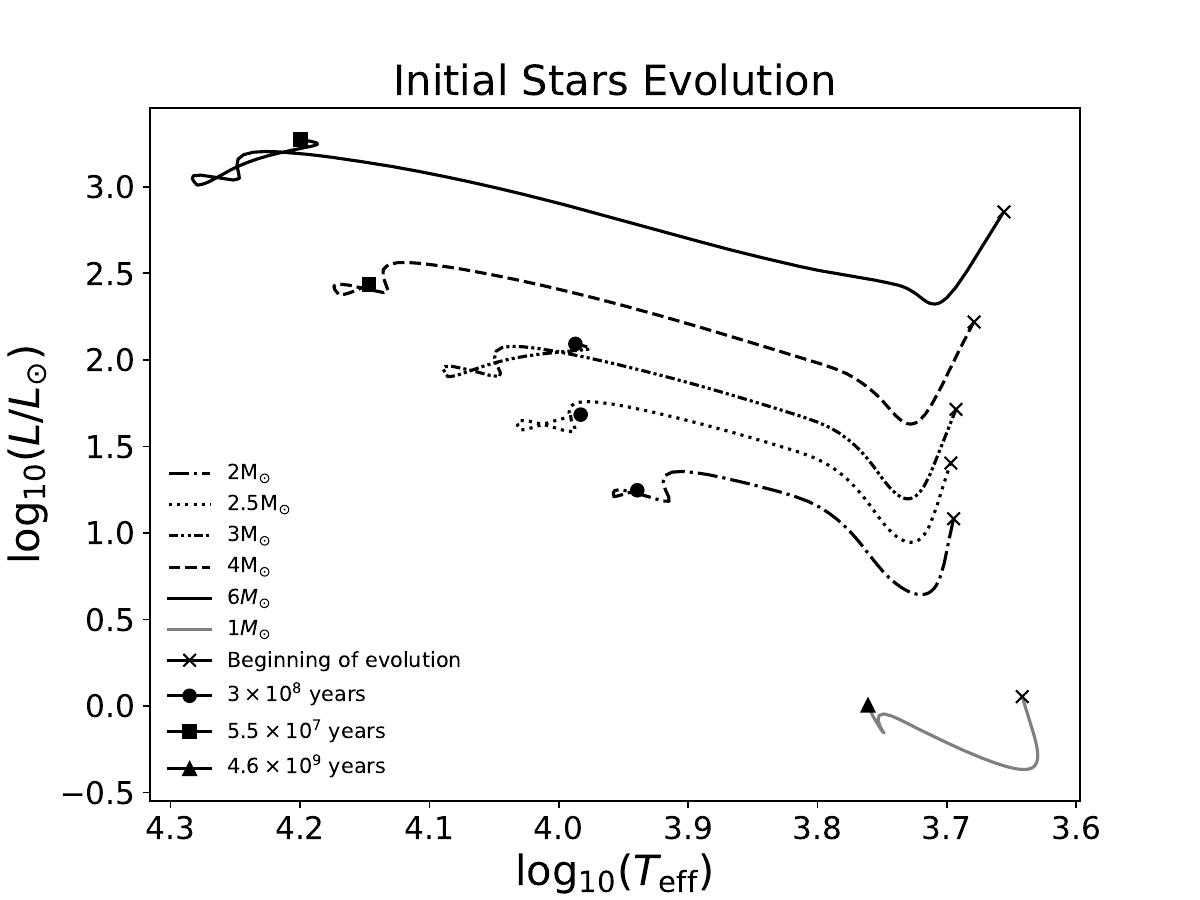}
        \caption{Evolutionary tracks of the parent stars. The squares denote a time of $5.5\times10^7$ years, corresponding to the age of a $6M_{\odot}$ TAMS star. This is the point at which the $4M_{\odot}$ and $6M_{\odot}$ stars are involved in the collision. The circles denote a time of $3\times10^8$ years, which corresponds with the age of a $3\,M_{\odot}$ TAMS star. This is the point at which the $3\,M_{\odot}$, $2.5M_{\odot}$, and $2\,M_{\odot}$ and stars are used in the collision. The $1M_{\odot}$ star is chosen at a time of $4.6\times10^9$ years, when its radius equals its time-weighted average main-sequence radius.}
        \label{fig:HR}
    \end{figure}

We model six stars to be used in the calculations throughout. The first three, a $2\,M_{\odot}$ MS, $2.5\,M_{\odot}$ MS, and $3\,M_{\odot}$ terminal age MS (TAMS) star, all roughly the same age of $3\times10^8$ years, are used in various collisions.
We define the TAMS as having a central hydrogen abundance $X\approx 0.01$.
These stars are chosen as they are slightly higher than the lowest mass required for significant CNO burning, and are therefore common stars to experience the CNO nitrogen enhancements.
We model collisions of these stars with a set of $3\,M_{\odot}$ + $2\,M_{\odot}$, $3\,M_{\odot}$ + $2.5\,M_{\odot}$, and $2.5\,M_{\odot}$ + $2\,M_{\odot}$ cases.
Additionally, a second set of stars, a $4\,M_{\odot}$ MS and a coeval $6\,M_{\odot}$ TAMS star are evolved for collisions between these two.
The last model considered is a $1\,M_{\odot}$ MS star with a radius of $1\,R_{\odot}$, its time-weighted average MS radius.
This star is interesting because it has not undergone significant CNO burning, but it can still strip substantial mass from the larger stars that it collides with \citep{FreitagBenz2005}.
Collisions between this $1\,M_{\odot}$ star and the other, intermediate-mass stars are modeled.
Detailed information regarding these stars, including age, radius, and abundances, can be found in Table \ref{table:parent_stars}, while their stellar evolutionary tracks are shown in Figure \ref{fig:HR}.

    \begin{deluxetable*}{ccc|ccccc|ccccc|cc}
        \label{table:parent_stars}
        \tabletypesize{\scriptsize}
        \tablecolumns{13}
        \tablewidth{0pt}
        \tablecaption{Parent Stars}
        \tablehead{&&& \multicolumn{5}{c|}{Central Abundances}  & \multicolumn{5}{c|}{Average Abundances} \\ \hline
            \colhead{$M$} & \colhead{$R$} & \colhead{age} &
            \colhead{H} & \colhead{He} & \colhead{C} & \colhead{N} & \colhead{O} &
            \colhead{H} & \colhead{He} & \colhead{C} & \colhead{N} & \colhead{O} & \colhead{$\rho_{\rm c}$} &
            \colhead{$P_{\rm c}$} \\ \hline
        $[M_{\odot}]$ &
        $[R_{\odot}]$ &
        [yr] &&&
        $\times10^{-5}$ &
        $\times10^{-3}$ &
        $\times10^{-4}$ &&&
        $\times10^{-3}$ &
        $\times10^{-3}$ &
        $\times10^{-3}$ &
        [g\,cm$^{-3}$] &
        [g\,cm$^{-1}$\,s$^{-2}$]
        }
        \startdata
            1 & $1.00$ & $4.6\times10^{9}$ & $0.344$ & $0.635$ & $2.16$ & $5.54$ & $87.7$ & $0.662$ & $0.317$ & $2.55$ & $2.06$ & $9.37$ & 151 & $2.33\times10^{17}$ \\ 
            2 & 1.85 & $3.0\times 10^8$ & 0.554 & 0.426 & $5.18$ & $8.64$ & $51.9$ & 0.675 & 0.304 & $2.00$ & $3.25$ & $8.76$ & 64.1 & $1.65\times10^{17}$ \\
            2.5 & 2.50 & $3.1\times 10^8$ & 0.382 & 0.599 & $7.82$ & $11.2$ & $21.9$ & 0.646 & 0.333 & $1.92$ & $3.85$ & $8.17$ & 49.9 & $1.21\times10^{17}$ \\
            3 & 3.93 & $3.1\times 10^8$ & 0.013 & 0.968 & $13.1$ & $12.6$ & $5.48$ & 0.601 & 0.379 & $1.89$ & $4.19$ & $7.82$ & 72.0 & $1.43\times10^{17}$ \\
            4 & 2.79 & $5.5\times10^7$ & 0.543 & 0.437 & $8.16$ & $10.2$ & $31.1$ & 0.666 & 0.314 & $1.87$ & $4.02$ & $8.04$ & 25.3 & $7.86\times10^{16}$ \\ 
            6 & 5.75 & $5.5\times10^7$ & $9.66\times10^{-3}$ & 0.971 & $17.2$ & $12.8$ & $3.29$ & 0.268 & 0.401 & $4.01$ & $4.78$ & $7.28$ & 31.1 & $7.81\times10^{16}$ \\ 
            \hline
            \multicolumn{15}{l}{
            \parbox{\linewidth}{\vspace{5pt}\textit{Notes:} Stellar properties and chemical abundances for the parent star models used in this study. Each star has an initial chemical abundance of $Y=0.28$ and $Z=0.02$.}
            \vspace{5pt}}\\
        \enddata
    \end{deluxetable*}

\subsection{{\tt StarSmasher}} \label{subsec:starsmasher}

We model the stellar collisions with {\tt StarSmasher} \citep{Gaburov+10,1991_Rasio_Thesis}, a Lagrangian smoothed particle hydrodynamics (SPH) code.\footnote{
{\tt StarSmasher} is publicly available at \url{https://github.com/jalombar/starsmasher}}
Each fluid particle $i$ is characterized by its mass $m_i$, position ${\bf r}_i$, velocity ${\bf v}_i$, and specific internal energy $u_i$.
Additionally, each particle is assigned a smoothing length $h_i$ that determines the local spatial resolution:
the Wendland C4 kernel \citep{Wendland1995PiecewisePP} provides the smoothing with a compact support of radius $2h_i$.
The computation of gravitational forces and energies involves a direct summation on NVIDIA graphics cards, following the approach outlined in \cite{Gaburov+10}.
This method provides an enhanced accuracy compared to tree-based methods, albeit at a slower pace.
For a comprehensive understanding of {\tt StarSmasher}'s implementation of these processes, along with guidelines for smoothing length determination and time-stepping, we refer to \cite{Kremer+22}.

\begin{figure*}
        \centering
        \includegraphics[width=0.494\linewidth]{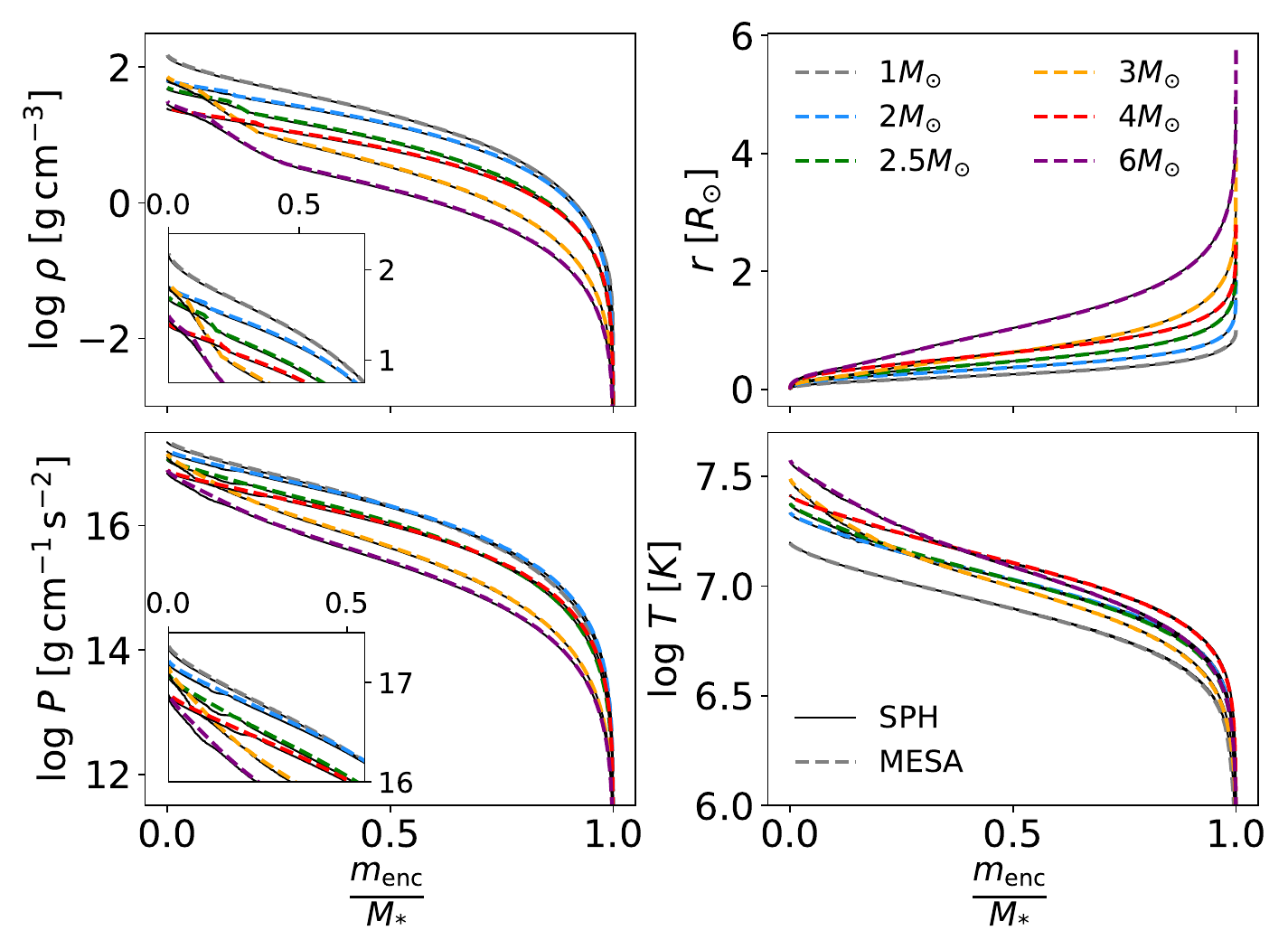}
        \includegraphics[width=0.495\linewidth]{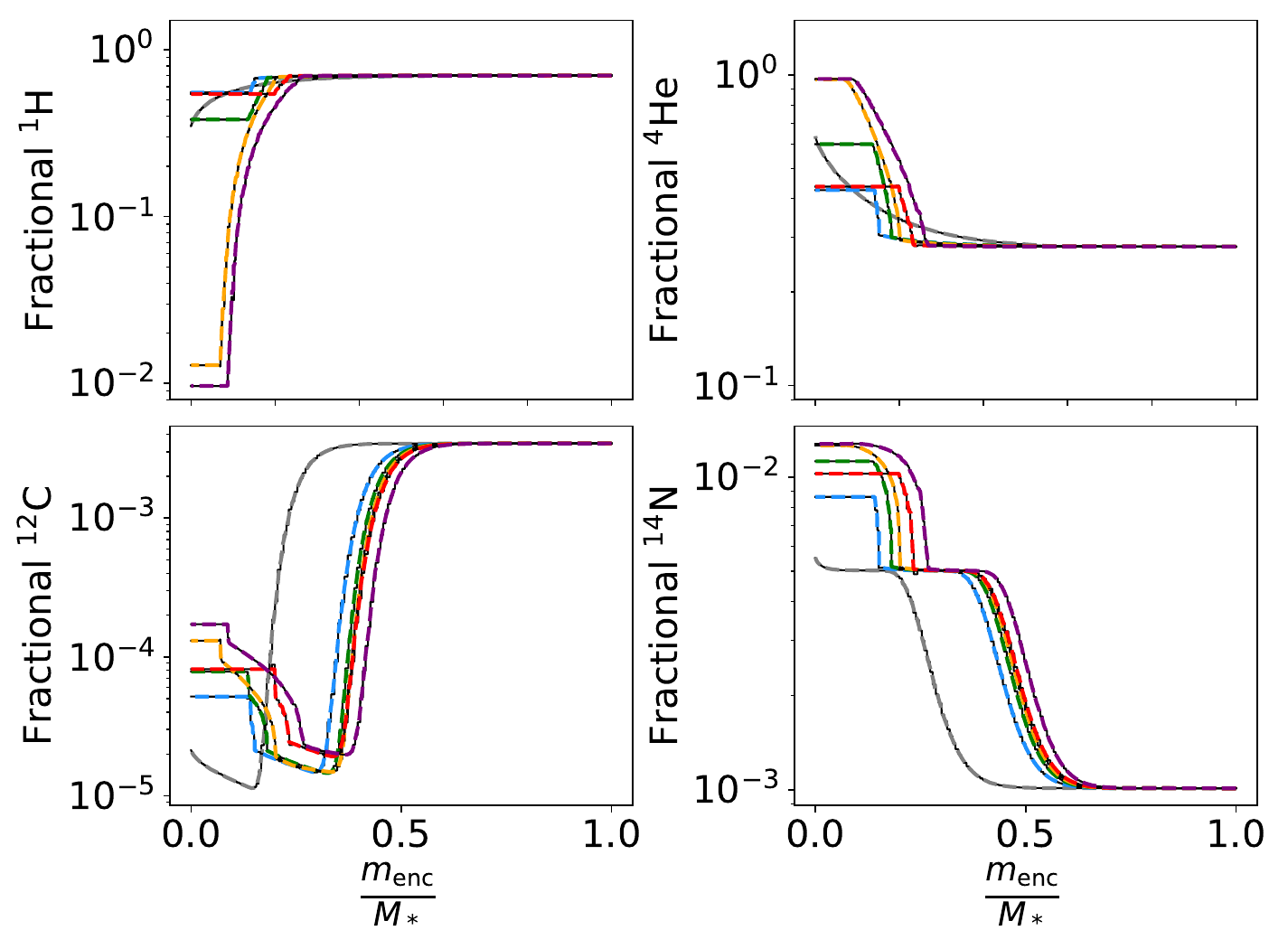}
        \caption{The SPH relaxation compared to the {\tt MESA} parent stars. The details of each star from Table \ref{table:parent_stars}. There is strong agreement between the quantities associated with each SPH particle (solid black lines) and those of the same location in the initial {\tt MESA} star (dashed colored lines on top of the SPH lines). Close to the centers, each star has a high nitrogen abundance relative to the carbon abundance at the same location, meaning that if substantial mass were to be stripped off, the overall average N/C abundance ratio would be higher than that of the parent star.
        }
        \label{fig:sph_mesa_relax}
    \end{figure*}

We convert the one-dimensional stellar evolution {\tt MESA} models into three-dimensional SPH models by placing the particles in a stretchy hexagonal close-packed (HCP) lattice as described in Appendix \ref{app:stretchy_hcp}.
After assigning initial parameters for the particles, the fluid is relaxed into hydrostatic equilibrium, as described in \cite{2006ApJ...640..441L,Gaburov+10}.
Our models are made of 250,000 particles each.
Figure~\ref{fig:sph_mesa_relax} demonstrates the agreement between the SPH and {\tt MESA} models of the stars.
The temperature, pressure, and specific internal energy values for each particle are evaluated via interpolation of a tabulated equation of state from {\tt MESA} for $Z=0.02$, the same metallicity used in Appendix~B of \cite{Paxton+18}.
This allows for a consistent EOS to be used between the original {\tt MESA} evolution, the collision calculations with SPH, and future evolution calculations again with {\tt MESA}.

Once the stars have obtained hydrostatic equilibrium through relaxation, they can be imported into the collision.
The stars are initially positioned along a hyperbolic trajectory and maintain an initial separation of $25\,R_{\odot}$, where tidal effects are negligible.
The distance of closest approach between the two stars, $r_{\rm p}$, is varied within $0\le r_{\rm p}/(R_1+R_2)\le0.5$, and the relative velocity at infinity, $v_{\infty}$ is varied within $0\le v_{\infty}\le 3\times10^4$\,km\,s$^{-1}$.
These $r_{\rm p}$ distances were chosen because any collision more off-axis is unlikely to strip significant mass from the stars.
The extreme range in $v_{\infty}$ values is chosen 
to include typical collision velocities within the Galactic Center.
Although shocks resulting from velocities this high are not well-resolved in SPH, they are treated well in terms of overall entropy production.
Additionally, the shocks are strongest in material that is ultimately unbound, and in this work, we focus on the structure and composition of the bound collision products.

To allow for a compositional analysis of the collision product, every particle is assigned a composition based on its radial position in the relaxed parent star.
The species used are the same ones tracked in the original {\tt MESA} evolution.
The composition of each fluid particle is then held constant for the duration of the collision.

Two types of instabilities could in principle cause mixing that we do not model: localized velocity shear instabilities \citep[e.g.,][]{2005AnRFM..37..329D} and more global instabilities like meridional circulation \citep[e.g.,][]{1972PASJ...24..509O,1995ApJ...440..789T}.  Although shear layers do exist in our simulations, they are primarily in fluid that ultimately becomes unbound by the collision.  Furthermore, the shear is between material from the envelopes of the parent stars, which have identical compositional makeups. As a result, this localized mixing would not significantly alter the overall composition profiles of the bound stellar product(s). In addition, although meridional circulation could drive more extensive mixing, it operates on the much longer Eddington-Sweet timescale and therefore would not act significantly on the timescale of our SPH simulation.

At the end of the collision, the final composition of each particle $i$ is determined by smoothing each value along the radius of the kernel, $2h_i$, and finding the compositional influence from neighboring particles.
Similarly, the specific angular momentum is smoothed across the kernel of the particle. The equations used are based on the form
\begin{equation}
    X_j=\frac{1}{\rho_j}\sum_k{X_km_kW_{jk}(h_j)},
\end{equation}
where $X$ is the value that is being smoothed and $W$ is the SPH kernel.
This is consistent with the SPH derivations outlined in \cite{Gaburov+10}.

\subsection{Analysis of Post-Collision Stellar Profiles}\label{subsec:sph_to_mesa}

Following the collision, we analyze the final structure of the remaining stars, beginning by determining the mass bound to each.
The iterative approach used to determine the final gravitationally bound masses $M_1$ and/or $M_2$ follows a similar methodology to that of \citet{2006ApJ...640..441L}.  However, based on the study by \citet{Nandez+14}, we use mechanical rather than total energy in the boundness criterion. Because the final states of the simulations within our parameter space region never result in a binary, there is no need to consider a common envelope. 

To create a one-dimensional model of the final product, significant time must pass between the time of the disruption and the time of analysis to ensure that most of the mass has approached an axisymmetric state.
In all of our simulations, we simulate at least 2 days of star time after a collision to meet this criterion.
At this point, we sphericalize the three-dimensional SPH data, sorting each particle into density bins, where the density is assumed to be decreasing from the center, acting as a radial profile.
This approach means that the center of the star is not the center of mass but rather the particle of highest density.
The mass-weighted averages of the position, velocity, density, specific internal energy, pressure, temperature, and composition are then evaluated for each bin, and continuous profiles of all tracked quantities are created.
A more resolved structure of the star is formed by linearly interpolating each of these values as a function of the enclosed mass fraction.

These profiles may then be used to evaluate the structure of the stars.
For example, the global average [N/C] data presented throughout this work are the ratios of total nitrogen to total carbon with respect to solar abundances, where the nitrogen and carbon abundances are integrated through the mass of the star.

    \begin{equation}
        10^{[{\rm N/C}]} =\frac{\int_0^{M_{\rm{tot}}}{\rm N}(m){\rm d}m}{\int_0^{M_{\rm{tot}}}{\rm C}(m){\rm d}m}\cdot\left(\frac{{\rm N}_{\odot}}{{\rm C}_{\odot}}\right)^{-1}
    \end{equation}

Following the collision, we construct two types of stellar profiles.
In the first type, we generate a hydrostatic equilibrium (HSE) model based on the entropy profile from the SPH calculation. In particular, we adopt the buoyancy $A$ profile from SPH as a proxy for entropy \citep{2008MNRAS.383L...5G} and solve the HSE equation for a slowly rotating star of the desired bound mass. We account for rotation by incorporating an effective gravitational acceleration that includes the lowest-order rotational correction based on the angular velocity $\omega$ profile \citep[e.g.,][]{1986A&A...157..329K}, approximated from the specific angular momentum $j$ and the mean radius $r$ as $\omega=3 j/(2r^2)$. Here we assume that the specific angular momentum profile is conserved from the SPH results to the HSE model. The models we consider (Cases 32 and 82) rotate well below break-up because most of the angular momentum is carried away in the ejecta of the collision. Consequently, the rotational corrections are minor and do not substantially affect the results.
Given the rapid structural evolution of a nascent stripped star immediately after its formation, our goal here is to generate a simplified one-dimensional model that preserves the thermally bloated size of the collisional product in a way that is not sensitive to the precise stopping time of the SPH calculation. Such an approach is preferred to extracting structure profiles directly from the SPH results since the outermost layers of a stripped star require considerable time to return to HSE after being ejected at speeds close to the escape speed.  When considering the future tidal disruption of such models, we refer to them as HSE models.

To generate the second type of one-dimensional model, we import the stripped star back into {\tt MESA} and evolve it until it returns to the main-sequence.
To accomplish this, we generate three files directly from the HSE SPH results following the {\tt MESA} relaxation method outlined in \cite{Paxton+18}: an entropy profile, a composition profile, and a specific angular momentum profile.
The entropy profile is calculated using the {\tt MESA} EOS table and a density-temperature pairing, where the density is obtained directly from {\tt StarSmasher} and the temperature is calculated using the same tabulated {\tt MESA} EOS used for the SPH calculations.
Because these {\tt MESA} models have thermally contracted to the main-sequence, these MS models are smaller than the corresponding HSE models.

Whether the HSE or {\tt MESA} MS model is more appropriate for a future TDE calculation depends on the post-collision trajectory of the stripped star.  If the stripped star is immediately headed into the loss cone of the SMBH, then the timescale before disruption
is some fraction of the orbital timescale and therefore very short.  In such a case, the thermally bloated HSE model is the more appropriate choice.  If instead the stripped star avoids an immediate TDE, then future gravitational scattering events could trigger the disruption.  In such a scenario, it is the two-body relaxation timescale that governs the time until the TDE, allowing the stripped star to contract to the main-sequence in the meantime \citep{Broggi+24}.
Future dynamical modeling is necessary to determine the relative likelihood of these two different scenarios.

\section{Results}\label{sec:results}

\begin{figure*}
        \centering
        
        \textbf{(A) Two Stripped Stars:} Case 32: $r_{\rm p}=(R_1+R_2)/8$, $v_{\infty}=5000$ km\,s$^{-1}$~~~~~~~~~~~~~~~~~~~~~~~~~~~~~~~~~~~~~~~~~~~~~~~~~~~~~~~~~~~~~~~~~~~~~~~~~~~~~
        
        \includegraphics[trim={0 3.56in 0 0},clip,width=0.9\linewidth]{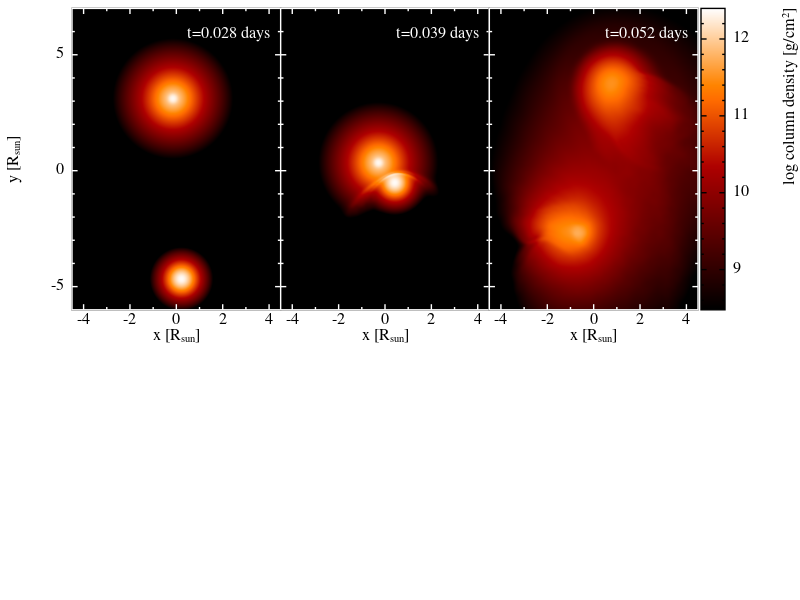} \\

    ~ \\
    
        \textbf{(B) One Stripped Star with no mass transfer:} Case 33: $r_{\rm p}=(R_1+R_2)/8$, $v_{\infty}=6000$ km\,s$^{-1}$~~~~~~~~~~~~~~~~~~~~~~~~~~~~~~~~~~~~~~~~~
    
        \includegraphics[trim={0 2.48in 0 0},clip,width=0.9\linewidth]{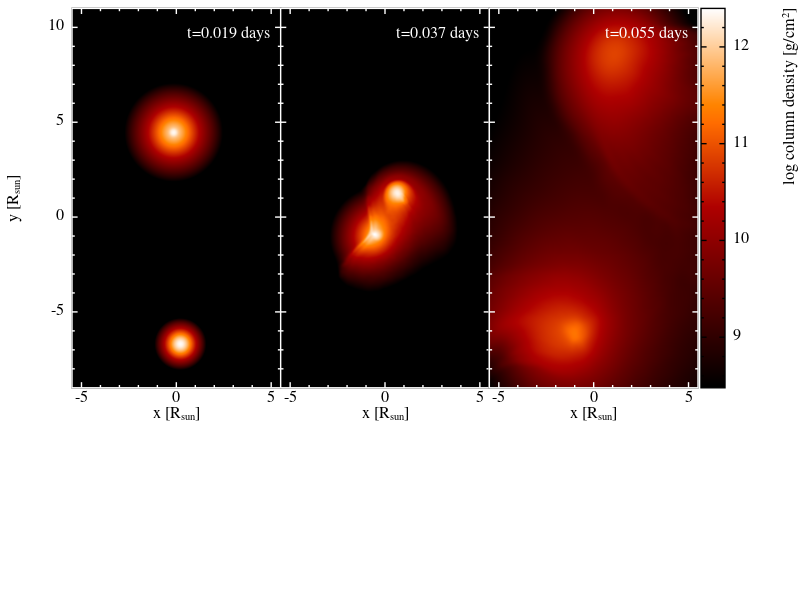} \\

~ \\
        
        \textbf{(C) One Stripped Star with mass transfer:} Case 6: $r_{\rm p}=0$, $v_{\infty}=3500$ km\,s$^{-1}$~~~~~~~~~~~~~~~~~~~~~~~~~~~~~~~~~~~~~~~~~~~~~~~~~~~~~~~~~~~~~~~~

        \includegraphics[trim={0 5.7in 0 0},clip,width=0.9\linewidth]{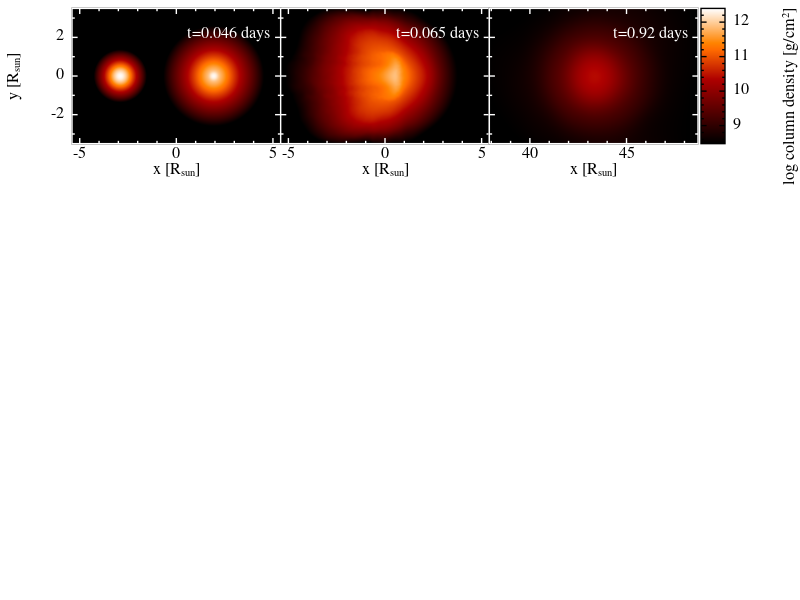}

        \caption{
        Various collisions of the $3\,M_{\odot}$ and $2\,M_{\odot}$ stars. \textbf{(A)} The top row illustrates Case 32, resulting in the creation of two stripped stars. \textbf{(B)} The second row of snapshots is for Case~33; although two separate overdense regions are seen in the final panel, the top one dissipates as time progresses, leaving just one stripped star. \textbf{(C)} The third row shows Case~6, in which one star remains with high mass contributed from both stars. Movies made from these collision calculations are available at \href{http://physics.allegheny.edu/jalombar/movies/}{http://physics.allegheny.edu/jalombar/movies/}.}
        \label{fig:collision_summary}
    \end{figure*}

    \begin{figure*}
        \includegraphics[width=\linewidth]{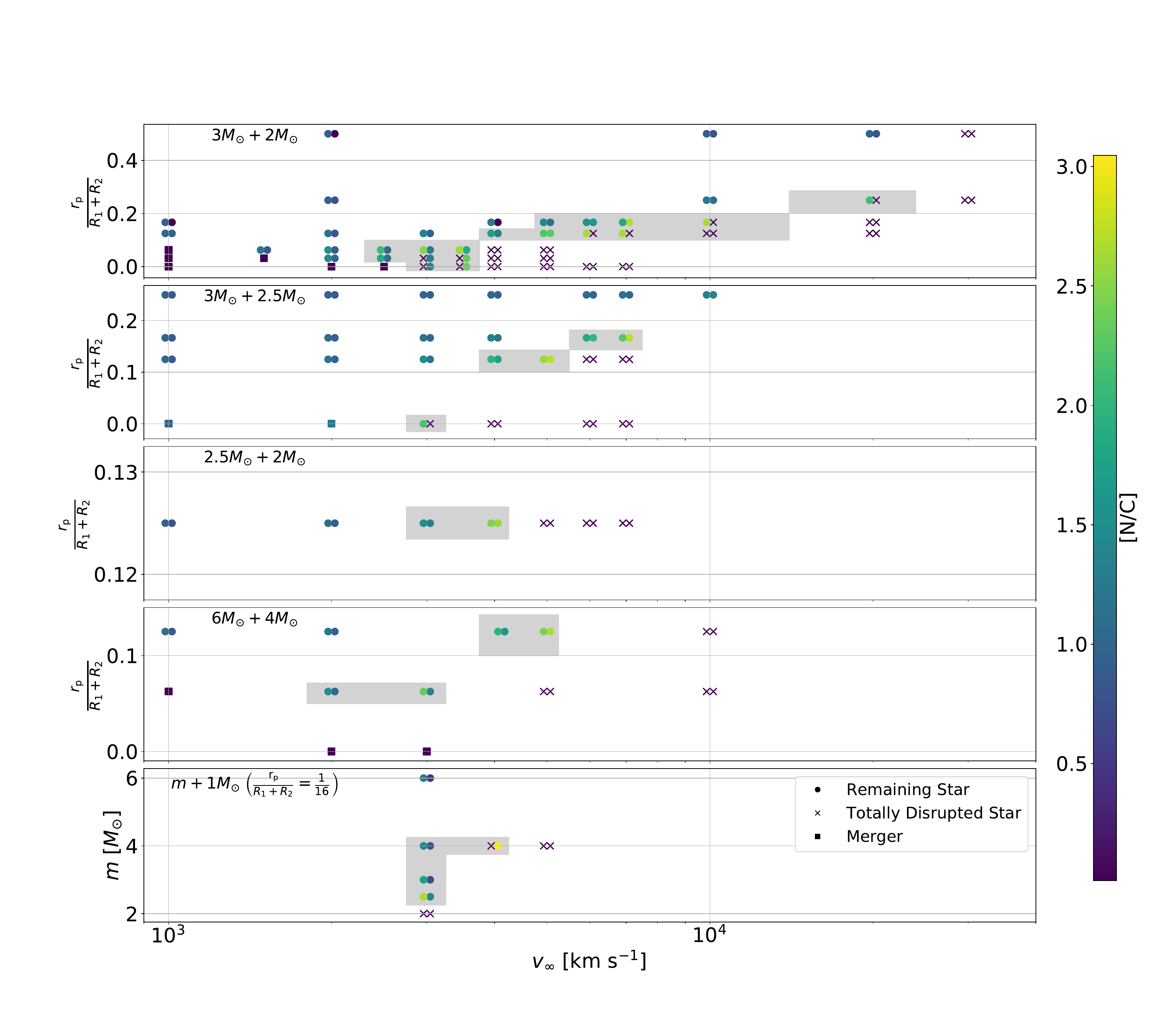}
        \caption{The $r_{\rm p}$ and $v_{\infty}$ data from Tables \ref{table:2_3_physical_table} and \ref{table:chem_table}, excluding the $v_{\infty}=0$ cases. The circles represent a single star remaining after the collision with the circles on the left corresponding to the product of star 1 in the collision and the circles on the right corresponding to the product of star 2 in the collision. A star is considered to be a product of Star 1 or Star 2 based on the product's most dense particle.
        The ``x" represents when a star has been completely destroyed. A square represents when the two stars result in a merger, defined as having one star remaining with a greater mass than the parent star.
        The color of each point corresponds to the global average [N/C] value of the star. The shaded regions show the regions in which there is an N/C greater than the constraints proposed by \citet{Yang17}. Stars that have an elevated [N/C] value have had much of their outer layers ejected during the collision and are plausible candidates for stripped star TDEs.}
        \label{fig:rp_v_vinf}
    \end{figure*}

A total of 106 simulations are presented in Tables \ref{table:2_3_physical_table} and \ref{table:chem_table}.
The average number of particles in each (bound) collision product is roughly 85,000, and the median is roughly 63,000.
Star 1 from Case 24 has only 811 particles remaining (the smallest particle number for any of our simulations), so a higher resolution collision was run (Case 25) between a 650,000 particle $3\,M_{\odot}$ star and the same 250,000 particle $2\,M_{\odot}$ star, yielding a star with 1899 particles.
This calculation gives similar results to the lower resolution collision.

We define a stripped star to be one for which its final bound mass is less than half of that of the corresponding parent star. In our results, stripped stars always resulted in an N/C with values ranging from $21.36 < 10^{[\rm N/C]} < 560$ (with an outlier of Case 104 with $10^{[\rm N/C]}=1118$). Based on this description, we identify four qualitative scenarios from our various collision simulations:

\textbf{1. Two stripped stars (2SS):}
In the first row of Figure \ref{fig:collision_summary}, we show Case 32. Here both stars survive, within final masses of $0.89M_{\odot}$ and $0.54M_{\odot}$ and average $10^{[{\rm N/C}]}$ values of 167 and 182, respectively. Neither of the stars experienced mass transfer during this collision.

\textbf{2. One stripped star (1SS):}
In the second row of Figure \ref{fig:collision_summary}, showing Case 33, only one star survives (with a mass of $0.62\,M_{\odot}$ and $10^{[\rm{N/C}]}$ of 442), with the second having been unbound by the high-velocity, off-axis impact. All of this collision product's mass comes from the $3\,M_{\odot}$ star.
However, some of these collision products retain high N/C values following the collision while also having experienced mass transfer between the two stars. An example of this, shown in the third row of Figure \ref{fig:collision_summary}, is Case 6.
The remaining star from Case 6 has a final mass of $1.07\,M_{\odot}$ and a $10^{\rm[N/C]}$ of 235. 17\% of the mass of the collision product comes from the $3\,M_{\odot}$ star, while the remaining $83\%$ of the mass comes from the $2\,M_{\odot}$ star.
Other examples of collisions yielding just one stripped star with elevated N/C include those that yield two products with only one retaining an elevated N/C abundance ratio (e.g., Cases 31, 92) and those that completely unbind the other star, similar to Case 32.

\textbf{3. No Stripped Stars (0SS):}
This result occurs when one or both of the stars survive the collision but do not experience significant enough mass loss to expose the high N/C material. This kind of outcome occurs from both grazing and lower-velocity collisions.
Grazing collisions are not direct enough to eject significant mass from the stars, whereas lower-velocity collisions typically result in mergers without ejecting the hydrogen-rich envelopes of either star.
There were not any mergers that yielded a high N/C quantity.
We distinguish mergers from collision products by defining mergers as having a greater mass than either of the original stars (or having a negative fractional mass loss).

\textbf{4. Complete Disruption (CD):}
This case is somewhat opposite to the previous one. Collisions that yield complete disruption are both direct enough and high enough velocity to completely eject all the mass from both stars, leaving no bound product behind.

In Figure \ref{fig:rp_v_vinf}, we show the outcome of all simulations listed in Table 2 (excluding $v_{\infty}=0\,\rm km\,s^{-1}$), with the top four panels showing $r_{\rm p}$ versus $v_{\infty}$ and the bottom panel showing $m$ versus $v_{\infty}$ for the cases involving the $1M_{\odot}$ star.
The various symbols denote different outcomes for the simulations, as described in the figure caption.
Colors denote the final [N/C] ratio of the collision products.
Figure \ref{fig:rp_v_vinf} demonstrates that, for the combination of the MS stars considered in this work, there is a small area in parameter space (shaded in the figure) in which stripped stars can be formed, requiring $v_{\infty}\gtrsim2000$\,km\,s$^{-1}$ with a strong $r_{\rm p}$ dependence. 
The $v_{\infty}$ values show remarkable agreement with the possible velocities in the Galactic Center outlined in Figure \ref{fig:tcoll}.
As the collisions become more off-axis at higher $r_{\rm{p}}$ values, the speeds required to strip the hydrogen-rich envelope from the star increase.
However, the same speeds required at more off-axis collisions would completely destroy both stars in a more head on collision.

\begin{figure}[t]
    \centering
    \includegraphics[width=\linewidth]{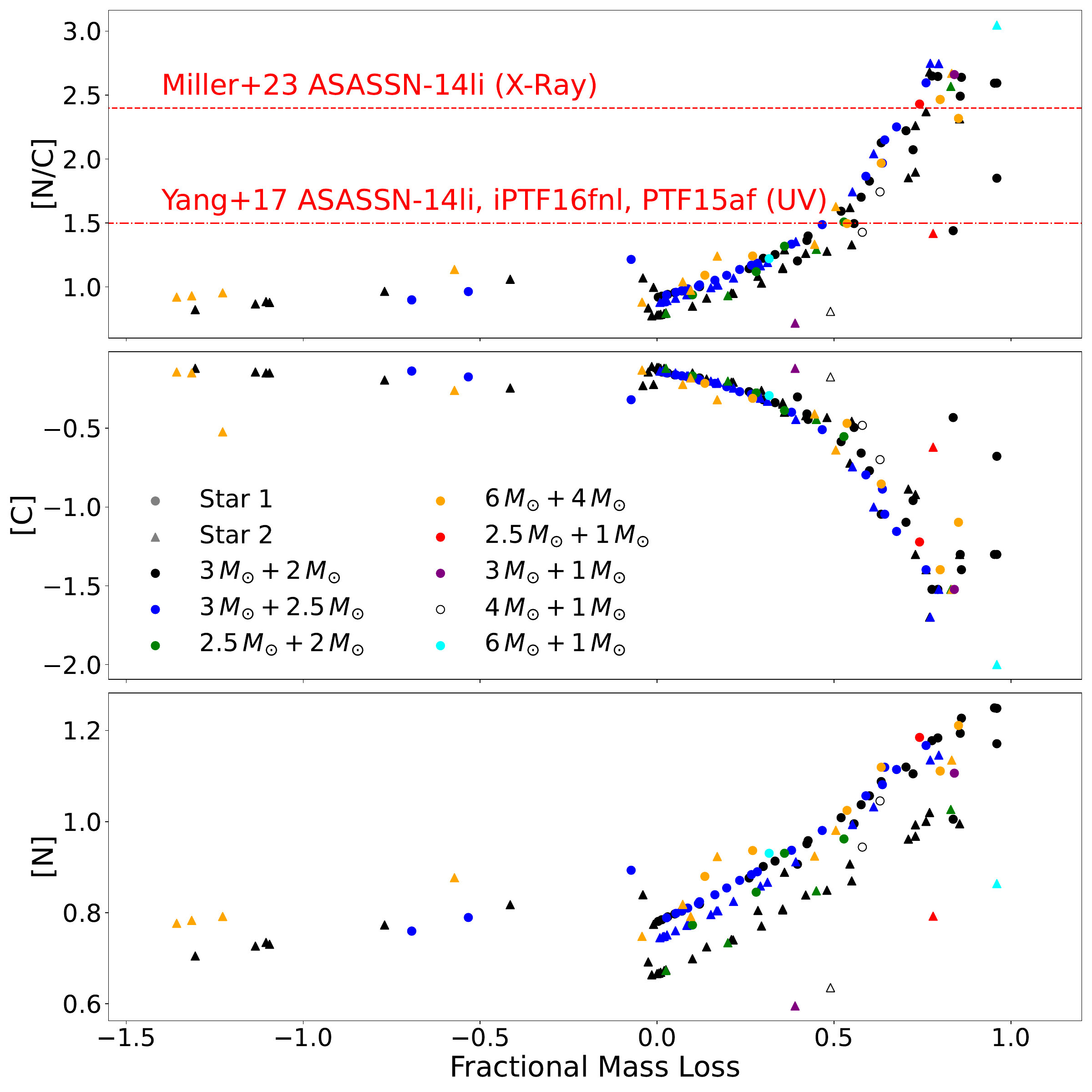}
    \caption{Abundance and abundance ratio information, relative to solar, vs the mass loss of the individual stars involved in the collision. Only the surviving products are shown. Data points with negative fractional mass loss correspond to merger products, which have a greater total mass than the original star. Although stripping the outer layers of the star begins to increase the [N/C] value, the highest [N/C] quantity is not necessarily observed when all or most of the mass is stripped. N/C peaks when roughly $90\%$ of the mass is ejected.
    The dashed, red horizontal lines give the lower limit of N/C based on various analyses of observed TDE spectra.
    }
    \label{fig:massloss_v_nc}
\end{figure}

The initial stellar profiles involved in the collision also influence the stripping of the outer envelope.
As the ratio of the mass of the two stars approaches 1, it becomes easier to strip mass from both stars, creating products with slightly higher N/C.
However, these stars are slightly more prone to total disruption at high velocities, limiting the range of velocities at which stripped stars may form.
This is especially noticeable when comparing the $3\,M_{\odot}+2.5\,M_{\odot}$ and $2.5\,M_{\odot}+2\,M_{\odot}$ cases to the $3\,M_{\odot}+2\,M_{\odot}$ cases.
For instance, at $r_{\rm p}=(R_1+R_2)/8$ for these three cases, both stars in the collision are completely destroyed at lower velocities for lower mass fraction values.
As the mass of the total system increases with a constant mass ratio, such as going from the $3\,M_{\odot}+2\,M_{\odot}$ to the $6\,M_{\odot}+4\,M_{\odot}$ cases, the fractional mass loss remains relatively consistent for the same $v_{\infty}$ and $r_{\rm p}/(R_1+R_2)$.
However, compositional differences between the corresponding stars (3 and 6 $M_{\odot}$ TAMS and 2 and 4 $M_{\odot}$ MS stars) lead to differences in the average N/C quantities despite the similar mass loss.

Additionally, it does not appear that significant CNO processing is necessarily required to create a stripped star with high N/C.
It was actually the remnant of the $1\,M_{\odot}$ star in Case 104 that yielded the highest N/C quantity in this study.
Although the nitrogen is not as significantly enhanced in the center of that star as in the higher mass stars, the low levels of carbon in the $1\,M_{\odot}$ star (see Figure~\ref{fig:sph_mesa_relax}) helps yield high N/C quantities.
This particular collision product has a mass of $0.04\,M_{\odot}$.
An object of this mass is consistent with the initial constraints of the progenitor of iPTF16fnl, which has estimates of the stellar mass as low as 0.03\,$M_{\odot}$ \citep{Blagorodnova+17}, although higher mass estimates up to 2.6\,$M_{\odot}$ also exist \citep{Mageshwaran23}.
This low-mass collision product, comprised of 1494 particles, appears stable in the simulation at late times.

Figure \ref{fig:massloss_v_nc} outlines the relationship between mass loss and N/C.
Generally, as more hydrogen- and carbon-rich mass, concentrated in the envelope, is ejected from the star, the total N/C value increases.
For very high mass loss, there is no longer a monotonically increasing N/C ratio for increasing fractional mass loss.
Instead, N/C begins to decrease very quickly.
This is expected based on Figure \ref{fig:sph_mesa_relax}, which shows that the greatest difference in nitrogen and carbon exists not at the center but rather at an enclosed mass fraction of $\sim20\%$.
    
    \startlongtable
    \begin{deluxetable*}{c|cccc|ccccccc|cc}
    \label{table:2_3_physical_table}
        \tabletypesize{\scriptsize}
        \tablecolumns{8}
        \tablewidth{0pt}
        \tablecaption{Post-Collision Stellar Products: Mechanical Properties}
        \tablehead{
            & \multicolumn{4}{c|}{Initial Conditions} & \multicolumn{7}{c|}{Final Star Configurations} & \multicolumn{2}{c}{Orbital Information} \\ \hline
            \colhead{Case No.} & \colhead{$M_{1_i}$} & \colhead{$M_{2_i}$} & \colhead{$\frac{r_p}{R_1+R_2}$} & \colhead{$v_{\infty,i}$} & 
            outcome &
            \colhead{$M_{1f}$} & \colhead{$M_{2f}$} &
            \colhead{$j_{\rm{rot},1}$} & \colhead{$j_{\rm{rot},2}$} & 
            \colhead{(\% Star 1)$_1$} &
            \colhead{(\% Star 1)$_2$} &
            \colhead{e} & \colhead{$v_{\infty,f}$} \\ \hline
            & [$M_{\odot}$] & [$M_{\odot}$] & & [km\,s$^{-1}$] & & [$M_{\odot}$] & [$M_{\odot}$] & [cm$^2$\,s$^{-1}$] & [cm$^2$\,s$^{-1}$] &&&& [km\,s$^{-1}$]
        }
        \startdata
            1 & 3.0 & 2.0 & 0 & 0 & 0SS & $-$ & 4.61 & $-$ & $7.78\times10^{14}$ & $-$ & 57.80 & $-$ & $-$ \\ 
            2 & 3.0 & 2.0 & 0 & 1000 & 0SS & $-$ & 4.19 & $-$ & $6.64\times10^{14}$ & $-$ & 54.90 & $-$ & $-$ \\ 
            3 & 3.0 & 2.0 & 0 & 2000 & 0SS & $-$ & 2.83 & $-$ & $2.14\times10^{15}$ & $-$ & 37.97 & $-$ & $-$ \\ 
            4 & 3.0 & 2.0 & 0 & 2500 & 0SS & $-$ & 2.08 & $-$ & $1.06\times10^{15}$ & $-$ & 24.56 & $-$ & $-$ \\ 
            5 & 3.0 & 2.0 & 0 & 3000 & 0SS & $-$ & 1.28 & $-$ & $1.45\times10^{15}$ & $-$ & 16.92 & $-$ & $-$ \\ 
            6 & 3.0 & 2.0 & 0 & 3500 & 1SS & $-$ & 0.48 & $-$ & $1.72\times10^{15}$ & $-$ & 5.14 & $-$ & $-$ \\ 
            7 & 3.0 & 2.0 & 0 & 4000 & CD & $-$ & $-$ & $-$ & $-$ & $-$ & $-$ & $-$ & $-$ \\ 
            8 & 3.0 & 2.0 & 0 & 5000 & CD & $-$ & $-$ & $-$ & $-$ & $-$ & $-$ & $-$ & $-$ \\ 
            9 & 3.0 & 2.0 & 0 & 6000 & CD & $-$ & $-$ & $-$ & $-$ & $-$ & $-$ & $-$ & $-$ \\ 
            10 & 3.0 & 2.0 & 0 & 7000 & CD & $-$ & $-$ & $-$ & $-$ & $-$ & $-$ & $-$ & $-$ \\ 
            11 & 3.0 & 2.0 & 1/32 & 1000 & 0SS & $-$ & 4.27 & $-$ & $9.65\times10^{17}$ & $-$ & 55.21 & $-$ & $-$ \\ 
            12 & 3.0 & 2.0 & 1/32 & 1500 & 0SS & $-$ & 3.54 & $-$ & $7.65\times10^{17}$ & $-$ & 48.26 & $-$ & $-$ \\ 
            13 & 3.0 & 2.0 & 1/32 & 2000 & 1SS & 0.49 & 2.02 & $1.06\times10^{17}$ & $2.41\times10^{17}$ & 80.96 & 21.45 & 1.043 & 262 \\ 
            14 & 3.0 & 2.0 & 1/32 & 2500 & 1SS & 0.12 & 1.43 & $8.87\times10^{15}$ & $1.90\times10^{17}$ & 87.60 & 9.01 & 4.698 & 695 \\ 
            15 & 3.0 & 2.0 & 1/32 & 3000 & 1SS & $-$ & 0.90 & $-$ & $1.24\times10^{17}$ & $-$ & 1.49 & 4.644 & 1035 \\ 
            16 & 3.0 & 2.0 & 1/32 & 3500 & 1SS & $-$ & 0.29 & $-$ & $2.68\times10^{16}$ & $-$ & 0.00 & $-$ & $-$ \\ 
            17 & 3.0 & 2.0 & 1/32 & 4000 & CD & $-$ & $-$ & $-$ & $-$ & $-$ & $-$ & $-$ & $-$ \\ 
            18 & 3.0 & 2.0 & 1/32 & 5000 & CD & $-$ & $-$ & $-$ & $-$ & $-$ & $-$ & $-$ & $-$ \\ 
            19 & 3.0 & 2.0 & 1/16 & 1000 & 0SS & $-$ & 4.21 & $-$ & $1.44\times10^{18}$ & $-$ & 54.33 & $-$ & $-$ \\ 
            20 & 3.0 & 2.0 & 1/16 & 1500 & 0SS & 1.81 & 2.05 & $4.08\times10^{17}$ & $3.12\times10^{17}$ & 94.31 & 13.56 & 1.041 & 472 \\ 
            21 & 3.0 & 2.0 & 1/16 & 2000 & 0SS & 1.33 & 1.72 & $2.03\times10^{17}$ & $2.05\times10^{17}$ & 95.43 & 4.90 & 2.57 & 988 \\ 
            22 & 3.0 & 2.0 & 1/16 & 2500 & 1SS & 0.83 & 1.41 & $8.38\times10^{16}$ & $1.40\times10^{17}$ & 97.19 & 0.59 & 4.743 & 1434 \\ 
            23 & 3.0 & 2.0 & 1/16 & 3000 & 1SS & 0.43 & 1.04 & $1.01\times10^{16}$ & $8.24\times10^{16}$ & 98.89 & 0.00 & 21.6 & 1871 \\ 
            24 & 3.0 & 2.0 & 1/16 & 3500 & 2SS & 0.12 & 0.54 & $1.34\times10^{16}$ & $2.80\times10^{16}$ & 100.00 & 0.00 & 94.35 & 2329 \\ 
            25* & 3.0 & 2.0 & 1/16 & 3500 & 2SS & 0.14 & 0.58 & $1.57\times10^{16}$ & $2.94\times10^{16}$ & 100.00 & 0.00 & 77.77 & 2309 \\ 
            26 & 3.0 & 2.0 & 1/16 & 4000 & CD & $-$ & $-$ & $-$ & $-$ & $-$ & $-$ & $-$ & $-$ \\ 
            27 & 3.0 & 2.0 & 1/16 & 5000 & CD & $-$ & $-$ & $-$ & $-$ & $-$ & $-$ & $-$ & $-$ \\ 
            28 & 3.0 & 2.0 & 1/8 & 1000 & 0SS & 2.64 & 2.03 & $3.81\times10^{17}$ & $2.56\times10^{17}$ & 97.80 & 5.99 & 1.175 & 495 \\ 
            29 & 3.0 & 2.0 & 1/8 & 2000 & 0SS & 2.22 & 1.80 & $1.28\times10^{17}$ & $1.07\times10^{17}$ & 99.47 & 0.86 & 1.22 & 1564 \\ 
            30 & 3.0 & 2.0 & 1/8 & 3000 & 0SS & 1.73 & 1.57 & $7.32\times10^{16}$ & $6.42\times10^{16}$ & 99.99 & 0.00 & 7.217 & 2516 \\ 
            31 & 3.0 & 2.0 & 1/8 & 4000 & 1SS & 1.27 & 1.16 & $5.18\times10^{16}$ & $3.83\times10^{16}$ & 100.00 & 0.00 & 14.84 & 3467 \\ 
            32 & 3.0 & 2.0 & 1/8 & 5000 & 2SS & 0.89 & 0.54 & $3.56\times10^{16}$ & $1.47\times10^{16}$ & 100.00 & 0.00 & 27.79 & 4426 \\ 
            33 & 3.0 & 2.0 & 1/8 & 6000 & 1SS & 0.62 & $-$ & $2.06\times10^{16}$ & $-$ & 100.00 & $-$ & $-$ & $-$ \\ 
            34 & 3.0 & 2.0 & 1/8 & 7000 & 1SS & 0.42 & $-$ & $1.21\times10^{16}$ & $-$ & 100.00 & $-$ & $-$ & $-$ \\ 
            35 & 3.0 & 2.0 & 1/8 & 10000 & CD & $-$ & $-$ & $-$ & $-$ & $-$ & $-$ & $-$ & $-$ \\ 
            36 & 3.0 & 2.0 & 1/8 & 20000 & CD & $-$ & $-$ & $-$ & $-$ & $-$ & $-$ & $-$ & $-$ \\ 
            37 & 3.0 & 2.0 & 1/6 & 1000 & 0SS & 2.79 & 1.99 & $2.84\times10^{17}$ & $1.71\times10^{17}$ & 98.86 & 2.64 & 1.396 & 711 \\ 
            38 & 3.0 & 2.0 & 1/6 & 4000 & 0SS & 2.00 & 1.58 & $6.13\times10^{16}$ & $3.85\times10^{16}$ & 100.00 & 0.00 & 16.92 & 3678 \\ 
            39 & 3.0 & 2.0 & 1/6 & 5000 & 0SS & 1.72 & 1.29 & $5.15\times10^{16}$ & $2.69\times10^{16}$ & 100.00 & 0.00 & 18.69 & 4653 \\ 
            40 & 3.0 & 2.0 & 1/6 & 6000 & 2SS & 1.44 & 0.91 & $4.27\times10^{16}$ & $1.85\times10^{16}$ & 100.00 & 0.00 & 19.98 & 5629 \\ 
            41 & 3.0 & 2.0 & 1/6 & 7000 & 2SS & 1.20 & 0.46 & $3.41\times10^{16}$ & $9.47\times10^{15}$ & 100.00 & 0.00 & 55.87 & 6608 \\ 
            42 & 3.0 & 2.0 & 1/6 & 10000 & 1SS & 0.67 & $-$ & $1.30\times10^{16}$ & $-$ & 100.00 & $-$ & $-$ & $-$ \\ 
            43 & 3.0 & 2.0 & 1/6 & 20000 & CD & $-$ & $-$ & $-$ & $-$ & $-$ & $-$ & $-$ & $-$ \\ 
            44 & 3.0 & 2.0 & 1/4 & 2000 & 0SS & 2.85 & 1.96 & $6.85\times10^{16}$ & $4.55\times10^{16}$ & 99.97 & 0.07 & 6.523 & 1912 \\ 
            45 & 3.0 & 2.0 & 1/4 & 10000 & 0SS & 2.10 & 1.29 & $3.81\times10^{16}$ & $1.38\times10^{16}$ & 100.00 & 0.00 & 142.4 & 9817 \\ 
            46 & 3.0 & 2.0 & 1/4 & 20000 & 1SS & 1.10 & $-$ & $2.51\times10^{16}$ & $-$ & 100.00 & $-$ & $-$ & $-$ \\ 
            47 & 3.0 & 2.0 & 1/4 & 30000 & CD & $-$ & $-$ & $-$ & $-$ & $-$ & $-$ & $-$ & $-$ \\ 
            48 & 3.0 & 2.0 & 1/2 & 2000 & 0SS & 2.99 & 2.00 & $1.06\times10^{16}$ & $4.44\times10^{15}$ & 100.00 & 0.00 & 12.8 & 1997 \\ 
            49 & 3.0 & 2.0 & 1/2 & 10000 & 0SS & 2.96 & 1.98 & $1.06\times10^{16}$ & $5.{15}\times10^{15}$ & 100.00 & 0.00 & 301.2 & 9992 \\ 
            50 & 3.0 & 2.0 & 1/2 & 20000 & 0SS & 2.91 & 1.95 & $1.29\times10^{16}$ & $6.86\times10^{15}$ & 100.00 & 0.00 & 1243 & 19985 \\ 
            51 & 3.0 & 2.0 & 1/2 & 30000 & CD & $-$ & $-$ & $-$ & $-$ & $-$ & $-$ & $-$ & $-$ \\ \hline 
            52 & 3.0 & 2.5 & 0 & 0 & 0SS & 5.08 & $-$ & $2.99\times10^{14}$ & $-$ & 52.97 & $-$ & $-$ & $-$ \\ 
            53 & 3.0 & 2.5 & 0 & 1000 & 0SS & 4.60 & $-$ & $2.44\times10^{16}$ & $-$ & 50.81 & $-$ & $-$ & $-$ \\ 
            54 & 3.0 & 2.5 & 0 & 2000 & 0SS & 3.22 & $-$ & $9.59\times10^{14}$ & $-$ & 38.82 & $-$ & $-$ & $-$ \\ 
            55 & 3.0 & 2.5 & 0 & 3000 & 1SS & 1.07 & $-$ & $1.34\times10^{15}$ & $-$ & 34.56 & $-$ & $-$ & $-$ \\ 
            56 & 3.0 & 2.5 & 0 & 4000 & CD & $-$ & $-$ & $-$ & $-$ & $-$ & $-$ & $-$ & $-$ \\ 
            57 & 3.0 & 2.5 & 0 & 6000 & CD & $-$ & $-$ & $-$ & $-$ & $-$ & $-$ & $-$ & $-$ \\ 
            58 & 3.0 & 2.5 & 0 & 7000 & CD & $-$ & $-$ & $-$ & $-$ & $-$ & $-$ & $-$ & $-$ \\ 
            59 & 3.0 & 2.5 & 1/8 & 1000 & 0SS & 2.65 & 2.46 & $4.67\times10^{17}$ & $3.47\times10^{17}$ & 96.40 & 5.79 & 1.009 & 467 \\ 
            60 & 3.0 & 2.5 & 1/8 & 2000 & 0SS & 2.15 & 2.12 & $1.41\times10^{17}$ & $1.35\times10^{17}$ & 99.52 & 0.82 & 1.367 & 1544 \\ 
            61 & 3.0 & 2.5 & 1/8 & 3000 & 0SS & 1.60 & 1.72 & $7.28\times10^{16}$ & $6.89\times10^{16}$ & 100.00 & 0.00 & 8.399 & 2503 \\ 
            62 & 3.0 & 2.5 & 1/8 & 4000 & 2SS & 1.09 & 1.12 & $4.35\times10^{16}$ & $3.29\times10^{16}$ & 100.00 & 0.00 & 1.756 & 3456 \\ 
            63 & 3.0 & 2.5 & 1/8 & 5000 & 2SS & 0.72 & 0.51 & $2.32\times10^{16}$ & $1.33\times10^{16}$ & 100.00 & 100.00 & 41.44 & 4411 \\ 
            64 & 3.0 & 2.5 & 1/8 & 6000 & CD & $-$ & $-$ & $-$ & $-$ & $-$ & $-$ & $-$ & $-$ \\ 
            65 & 3.0 & 2.5 & 1/8 & 7000 & CD & $-$ & $-$ & $-$ & $-$ & $-$ & $-$ & $-$ & $-$ \\ 
            66 & 3.0 & 2.5 & 1/6 & 1000 & 0SS & 2.79 & 2.45 & $3.36\times10^{17}$ & $2.44\times10^{17}$ & 98.32 & 2.54 & 1.798 & 704 \\ 
            67 & 3.0 & 2.5 & 1/6 & 2000 & 0SS & 2.51 & 2.28 & $1.{17}\times10^{17}$ & $1.06\times10^{17}$ & 99.84 & 0.37 & 2.487 & 1739 \\ 
            68 & 3.0 & 2.5 & 1/6 & 3000 & 0SS & 2.20 & 2.07 & $7.54\times10^{16}$ & $6.42\times10^{16}$ & 100.00 & 0.00 & 5.48 & 2711 \\ 
            69 & 3.0 & 2.5 & 1/6 & 4000 & 0SS & 1.86 & 1.77 & $5.90\times10^{16}$ & $4.37\times10^{16}$ & 100.00 & 0.00 & 7.621 & 3680 \\ 
            70 & 3.0 & 2.5 & 1/6 & 6000 & 2SS & 1.23 & 0.97 & $3.85\times10^{16}$ & $2.08\times10^{16}$ & 100.00 & 0.00 & 29.73 & 5627 \\ 
            71 & 3.0 & 2.5 & 1/6 & 7000 & 2SS & 0.97 & 0.57 & $3.00\times10^{16}$ & $1.35\times10^{16}$ & 100.00 & 0.00 & 31.09 & 6603 \\ 
            72 & 3.0 & 2.5 & 1/4 & 1000 & 0SS & 2.92 & 2.48 & $1.81\times10^{17}$ & $1.20\times10^{17}$ & 99.66 & 0.65 & 1.994 & 898 \\ 
            73 & 3.0 & 2.5 & 1/4 & 2000 & 0SS & 2.84 & 2.43 & $7.52\times10^{16}$ & $6.21\times10^{16}$ & 99.97 & 0.05 & 6.984 & 1913 \\ 
            74 & 3.0 & 2.5 & 1/4 & 3000 & 0SS & 2.74 & 2.37 & $5.74\times10^{16}$ & $4.74\times10^{16}$ & 100.00 & 0.00 & 13.96 & 2902 \\ 
            75 & 3.0 & 2.5 & 1/4 & 4000 & 0SS & 2.64 & 2.29 & $5.{16}\times10^{16}$ & $4.09\times10^{16}$ & 100.00 & 0.00 & 26.48 & 3890 \\ 
            76 & 3.0 & 2.5 & 1/4 & 6000 & 0SS & 2.41 & 2.08 & $4.53\times10^{16}$ & $3.25\times10^{16}$ & 100.00 & 0.00 & 38.59 & 5863 \\ 
            77 & 3.0 & 2.5 & 1/4 & 7000 & 0SS & 2.30 & 1.96 & $4.24\times10^{16}$ & $2.85\times10^{16}$ & 100.00 & 0.00 & 72.84 & 6852 \\ 
            78 & 3.0 & 2.5 & 1/4 & 10000 & 0SS & 1.92 & 1.52 & $3.73\times10^{16}$ & $2.07\times10^{16}$ & 100.00 & 0.00 & 180.9 & 9821 \\ \hline 
            79 & 2.5 & 2.0 & 1/8 & 1000 & 0SS & 2.25 & 1.95 & $4.03\times10^{17}$ & $3.41\times10^{17}$ & 95.44 & 7.15 & 1.025 & 253 \\ 
            80 & 2.5 & 2.0 & 1/8 & 2000 & 0SS & 1.80 & 1.60 & $1.12\times10^{17}$ & $1.{17}\times10^{17}$ & 99.38 & 1.07 & 1.495 & 1429 \\ 
            81 & 2.5 & 2.0 & 1/8 & 3000 & 1SS & 1.18 & 1.10 & $4.42\times10^{16}$ & $4.90\times10^{16}$ & 100.00 & 0.00 & 19.57 & 2386 \\ 
            82 & 2.5 & 2.0 & 1/8 & 4000 & 2SS & 0.50 & 0.34 & $2.03\times10^{16}$ & $1.07\times10^{16}$ & 100.00 & 0.00 & 69.55 & 3323 \\ 
            83 & 2.5 & 2.0 & 1/8 & 5000 & CD & $-$ & $-$ & $-$ & $-$ & $-$ & $-$ & $-$ & $-$ \\ 
            84 & 2.5 & 2.0 & 1/8 & 6000 & CD & $-$ & $-$ & $-$ & $-$ & $-$ & $-$ & $-$ & $-$ \\ 
            85 & 2.5 & 2.0 & 1/8 & 7000 & CD & $-$ & $-$ & $-$ & $-$ & $-$ & $-$ & $-$ & $-$ \\ \hline 
            86 & 6.0 & 4.0 & 0 & 0 & 0SS & $-$ & 9.26 & $-$ & $1.63\times10^{15}$ & $-$ & 57.90 & $-$ & $-$ \\ 
            87 & 6.0 & 4.0 & 0 & 2000 & 0SS & $-$ & 6.29 & $-$ & $4.08\times10^{15}$ & $-$ & 42.81 & $-$ & $-$ \\ 
            88 & 6.0 & 4.0 & 0 & 3000 & 0SS & $-$ & 3.32 & $-$ & $1.97\times10^{15}$ & $-$ & 17.80 & $-$ & $-$ \\ 
            89 & 6.0 & 4.0 & 1/16 & 500 & 0SS & $-$ & 9.43 & $-$ & $2.35\times10^{18}$ & $-$ & 0.00 & $-$ & $-$ \\ 
            90 & 6.0 & 4.0 & 1/16 & 1000 & 0SS & $-$ & 8.91 & $-$ & $2.41\times10^{18}$ & $-$ & 0.00 & $-$ & 56.36 \\ 
            91 & 6.0 & 4.0 & 1/16 & 2000 & 1SS & 2.78 & 3.71 & $5.31\times10^{17}$ & $4.63\times10^{17}$ & 93.73 & 9.96 & 1.288 & 763 \\ 
            92 & 6.0 & 4.0 & 1/16 & 3000 & 1SS & 0.89 & 2.22 & $1.79\times10^{16}$ & $2.22\times10^{17}$ & 97.75 & 0.00 & 13.9 & 1649 \\ 
            93 & 6.0 & 4.0 & 1/16 & 5000 & CD & $-$ & $-$ & $-$ & $-$ & $-$ & $-$ & $-$ & $-$ \\ 
            94 & 6.0 & 4.0 & 1/16 & 10000 & CD & $-$ & $-$ & $-$ & $-$ & $-$ & $-$ & $-$ & $-$ \\ 
            95 & 6.0 & 4.0 & 1/8 & 1000 & 0SS & 5.19 & 4.17 & $9.28\times10^{17}$ & $6.32\times10^{17}$ & 97.58 & 91.36 & 1.003 & 81.3 \\ 
            96 & 6.0 & 4.0 & 1/8 & 2000 & 0SS & 4.38 & 3.62 & $2.91\times10^{17}$ & $2.46\times10^{17}$ & 99.16 & 98.63 & 2.814 & 1425 \\ 
            97 & 6.0 & 4.0 & 1/8 & 4117 & 2SS & 2.20 & 1.98 & $9.68\times10^{16}$ & $6.65\times10^{16}$ & 100.00 & 0.00 & 18.26 & 3439 \\ 
            98 & 6.0 & 4.0 & 1/8 & 5000 & 2SS & 1.55 & 0.67 & $6.65\times10^{16}$ & $1.87\times10^{16}$ & 100.00 & 0.00 & 24.64 & 4274 \\ 
            99 & 6.0 & 4.0 & 1/8 & 10000 & CD & $-$ & $-$ & $-$ & $-$ & $-$ & $-$ & $-$ & $-$ \\ \hline 
            100 & 2.0 & 1.0 & 1/16 & 3000 & CD & $-$ & $-$ & $-$ & $-$ & $-$ & $-$ & $-$ & $-$ \\ \hline 
            101 & 2.5 & 1.0 & 1/16 & 3000 & 2SS &  0.40 & 0.22 & $6.84\times10^{15}$ & $1.81\times10^{16}$ & 99.90 & 0.00 & 33.12 & 1654 \\ \hline 
            102 & 3.0 & 1.0 & 1/16 & 3000 & 1SS & 1.11 & 0.61 & $5.10\times10^{16}$ & $5.68\times10^{16}$ & 98.68 & 0.00 & 19.37 & 1961 \\ \hline 
            103 & 4.0 & 1.0 & 1/16 & 3000 & 1SS & 1.68 & 0.51 & $2.66\times10^{16}$ & $8.32\times10^{16}$ & 99.41 & 0.00 & 8.99 & 1522 \\ 
            104 & 4.0 & 1.0 & 1/16 & 4000 & 1SS & $-$ & 0.04 & $-$ & $2.30\times10^{15}$ & $-$ & 0.00 & 1028 & 2380 \\ 
            105 & 4.0 & 1.0 & 1/16 & 5000 & CD & $-$ & $-$ & $-$ & $-$ & $-$ & $-$ & $-$ & $-$ \\ \hline 
            106 & 6.0 & 1.0 & 1/16 & 3000 & 0SS & 4.10 & 0.88 & $1.12\times10^{17}$ & $8.17\times10^{16}$ & 99.51 & 0.00 & 3.787 & 1954 \\ 
            \hline
            \multicolumn{14}{l}{
            \parbox{\linewidth}{\vspace{5pt}\textit{Notes:} 
Column 1 gives the case number, with an * marking our high-resolution simulation. Columns 2 and 3 give the masses of our parent stars.  Columns 4 and 5 give the initial orbital parameters, namely the periapsis separation normalized to the sum of the stellar radii and the velocity at infinity, respectively. Column 6 gives the qualitative outcome: 0SS = zero stripped stars, 1SS = one stripped star, 2SS = two stripped stars, CD = complete disruption. Columns 7 and 8 give the masses of the final bound components if they exist.  Columns 9 and 10 give the average specific angular momentum of each final bound component.  Columns 11 and 12 give the percentage of each component that originated in parent star 1 (the more massive parent).  Columns 13 and 14 give final orbital parameters for cases when two bound components remain, namely the eccentricity and relative velocity at infinity, respectively.}
            \vspace{5pt}}\\
        \enddata
    \end{deluxetable*}

    \startlongtable
    \begin{deluxetable*}{c|cc|cccccccccc}
    \label{table:chem_table}
        \tabletypesize{\scriptsize}
        \tablecolumns{13}
        \tablewidth{0pt}
        \tablecaption{Post-Collision Stellar Products: Average Chemical Abundances Relative to Solar. Fully disrupted cases are not included}
        \tablehead{
            \colhead{Case No.} &
            \colhead{$10^{[\rm{N/C}]_1}$} & 
            \colhead{$10^{[\rm{N/C}]_2}$} & 
            \colhead{$10^{[^1\rm{H}]_1}$} & 
            \colhead{$10^{[^1\rm{H}]_2}$} & 
            \colhead{$10^{[^4\rm{He}]_1}$} & 
            \colhead{$10^{[^4\rm{He}]_2}$} & 
            \colhead{$10^{[^{12}\rm{C}]_1}$} & 
            \colhead{$10^{[^{12}\rm{C}]_2}$} & 
            \colhead{$10^{[^{14}\rm{N}]_1}$} & 
            \colhead{$10^{[^{14}\rm{N}]_2}$} & 
            \colhead{$10^{[^{16}\rm{O}]_1}$} & 
            \colhead{$10^{[^{16}\rm{O}]_2}$}
        }
        \startdata
            1 & $-$ & 6.64 & $-$ & 0.87 & $-$ & 1.36 & $-$ & 0.76 & $-$ & 5.07 & $-$ & 1.45 \\ 
            2 & $-$ & 7.56 & $-$ & 0.86 & $-$ & 1.38 & $-$ & 0.71 & $-$ & 5.38 & $-$ & 1.43 \\ 
            3 & $-$ & 11.49 & $-$ & 0.83 & $-$ & 1.49 & $-$ & 0.57 & $-$ & 6.57 & $-$ & 1.35 \\ 
            4 & $-$ & 11.76 & $-$ & 0.80 & $-$ & 1.57 & $-$ & 0.59 & $-$ & 6.91 & $-$ & 1.30 \\ 
            5 & $-$ & 19.50 & $-$ & 0.80 & $-$ & 1.58 & $-$ & 0.40 & $-$ & 7.74 & $-$ & 1.29 \\ 
            6 & $-$ & 234.51 & $-$ & 0.80 & $-$ & 1.57 & $-$ & 0.04 & $-$ & 10.01 & $-$ & 1.19 \\ 
            11 & $-$ & 7.36 & $-$ & 0.86 & $-$ & 1.38 & $-$ & 0.72 & $-$ & 5.33 & $-$ & 1.43 \\ 
            12 & $-$ & 9.23 & $-$ & 0.85 & $-$ & 1.43 & $-$ & 0.64 & $-$ & 5.93 & $-$ & 1.40 \\ 
            13 & 27.53 & 9.92 & 0.61 & 0.87 & 2.17 & 1.36 & 0.37 & 0.60 & 10.12 & 5.95 & 0.98 & 1.42 \\ 
            14 & 70.80 & 12.10 & 0.27 & 0.87 & 3.18 & 1.36 & 0.21 & 0.53 & 14.82 & 6.38 & 0.43 & 1.41 \\ 
            15 & $-$ & 21.36 & $-$ & 0.87 & $-$ & 1.38 & $-$ & 0.35 & $-$ & 7.41 & $-$ & 1.37 \\ 
            16 & $-$ & 206.12 & $-$ & 0.82 & $-$ & 1.50 & $-$ & 0.05 & $-$ & 9.89 & $-$ & 1.20 \\ 
            19 & $-$ & 7.68 & $-$ & 0.86 & $-$ & 1.38 & $-$ & 0.71 & $-$ & 5.43 & $-$ & 1.43 \\ 
            20 & 15.97 & 6.84 & 0.73 & 0.90 & 1.80 & 1.26 & 0.50 & 0.72 & 8.06 & 4.92 & 1.19 & 1.49 \\ 
            21 & 31.31 & 8.17 & 0.65 & 0.90 & 2.03 & 1.27 & 0.32 & 0.65 & 9.89 & 5.31 & 1.04 & 1.48 \\ 
            22 & 118.17 & 10.71 & 0.51 & 0.89 & 2.48 & 1.29 & 0.11 & 0.55 & 12.73 & 5.90 & 0.78 & 1.46 \\ 
            23 & 310.03 & 19.00 & 0.28 & 0.88 & 3.15 & 1.34 & 0.05 & 0.37 & 15.62 & 7.07 & 0.41 & 1.40 \\ 
            24 & 392.31 & 79.03 & 0.07 & 0.84 & 3.79 & 1.47 & 0.05 & 0.12 & 17.72 & 9.29 & 0.12 & 1.24 \\ 
            25 & 391.72 & 71.41 & 0.07 & 0.84 & 3.80 & 1.46 & 0.05 & 0.13 & 17.76 & 9.15 & 0.12 & 1.26 \\ 
            28 & 10.04 & 5.94 & 0.79 & 0.91 & 1.62 & 1.25 & 0.66 & 0.78 & 6.59 & 4.61 & 1.30 & 1.50 \\ 
            29 & 13.91 & 7.07 & 0.76 & 0.90 & 1.71 & 1.26 & 0.54 & 0.71 & 7.52 & 5.00 & 1.24 & 1.49 \\ 
            30 & 23.09 & 8.91 & 0.70 & 0.90 & 1.87 & 1.28 & 0.39 & 0.62 & 8.94 & 5.50 & 1.13 & 1.47 \\ 
            31 & 50.30 & 18.30 & 0.62 & 0.88 & 2.13 & 1.32 & 0.22 & 0.38 & 10.89 & 6.90 & 0.97 & 1.42 \\ 
            32 & 166.52 & 182.35 & 0.49 & 0.82 & 2.52 & 1.50 & 0.08 & 0.05 & 13.17 & 9.84 & 0.73 & 1.21 \\ 
            33 & 442.10 & $-$ & 0.34 & $-$ & 2.99 & $-$ & 0.03 & $-$ & 15.26 & $-$ & 0.47 & $-$ \\ 
            34 & 435.13 & $-$ & 0.19 & $-$ & 3.45 & $-$ & 0.04 & $-$ & 16.86 & $-$ & 0.25 & $-$ \\ 
            37 & 9.30 & 6.06 & 0.79 & 0.91 & 1.59 & 1.25 & 0.68 & 0.77 & 6.36 & 4.65 & 1.32 & 1.50 \\ 
            38 & 17.92 & 8.95 & 0.73 & 0.90 & 1.78 & 1.28 & 0.46 & 0.62 & 8.19 & 5.51 & 1.20 & 1.47 \\ 
            39 & 25.04 & 13.92 & 0.70 & 0.89 & 1.88 & 1.30 & 0.36 & 0.46 & 9.08 & 6.40 & 1.13 & 1.44 \\ 
            40 & 39.08 & 41.70 & 0.65 & 0.87 & 2.02 & 1.37 & 0.26 & 0.19 & 10.20 & 8.07 & 1.03 & 1.37 \\ 
            41 & 67.22 & 478.64 & 0.60 & 0.81 & 2.20 & 1.56 & 0.17 & 0.02 & 11.39 & 10.47 & 0.92 & 1.14 \\ 
            42 & 446.05 & $-$ & 0.36 & $-$ & 2.92 & $-$ & 0.03 & $-$ & 15.05 & $-$ & 0.50 & $-$ \\ 
            44 & 9.04 & 6.20 & 0.80 & 0.91 & 1.59 & 1.25 & 0.69 & 0.76 & 6.27 & 4.71 & 1.32 & 1.50 \\ 
            45 & 16.75 & 14.17 & 0.74 & 0.89 & 1.75 & 1.31 & 0.48 & 0.45 & 7.97 & 6.43 & 1.22 & 1.44 \\ 
            46 & 133.99 & $-$ & 0.56 & $-$ & 2.30 & $-$ & 0.09 & $-$ & 12.24 & $-$ & 0.85 & $-$ \\ 
            48 & 8.32 & 6.03 & 0.80 & 0.91 & 1.56 & 1.25 & 0.73 & 0.77 & 6.04 & 4.64 & 1.34 & 1.50 \\ 
            49 & 8.47 & 6.11 & 0.80 & 0.91 & 1.57 & 1.25 & 0.72 & 0.76 & 6.09 & 4.67 & 1.33 & 1.50 \\ 
            50 & 8.74 & 6.27 & 0.80 & 0.91 & 1.58 & 1.25 & 0.71 & 0.75 & 6.18 & 4.73 & 1.33 & 1.50 \\ \hline 
            52 & 7.92 & $-$ & 0.84 & $-$ & 1.45 & $-$ & 0.73 & $-$ & 5.75 & $-$ & 1.38 & $-$ \\ 
            53 & 9.20 & $-$ & 0.83 & $-$ & 1.48 & $-$ & 0.67 & $-$ & 6.16 & $-$ & 1.35 & $-$ \\ 
            54 & 16.41 & $-$ & 0.78 & $-$ & 1.63 & $-$ & 0.48 & $-$ & 7.82 & $-$ & 1.24 & $-$ \\ 
            55 & 141.12 & $-$ & 0.54 & $-$ & 2.37 & $-$ & 0.09 & $-$ & 13.16 & $-$ & 0.73 & $-$ \\ 
            59 & 10.14 & 7.66 & 0.79 & 0.87 & 1.62 & 1.37 & 0.65 & 0.73 & 6.61 & 5.59 & 1.30 & 1.40 \\ 
            60 & 15.33 & 9.86 & 0.75 & 0.85 & 1.73 & 1.41 & 0.51 & 0.63 & 7.76 & 6.25 & 1.23 & 1.36 \\ 
            61 & 30.69 & 15.53 & 0.68 & 0.83 & 1.93 & 1.47 & 0.31 & 0.47 & 9.56 & 7.36 & 1.09 & 1.30 \\ 
            62 & 92.88 & 55.42 & 0.56 & 0.78 & 2.30 & 1.65 & 0.13 & 0.18 & 12.05 & 9.85 & 0.86 & 1.13 \\ 
            63 & 394.15 & 556.79 & 0.39 & 0.61 & 2.83 & 2.15 & 0.04 & 0.03 & 14.69 & 13.99 & 0.55 & 0.65 \\ 
            66 & 9.35 & 7.70 & 0.79 & 0.87 & 1.59 & 1.37 & 0.68 & 0.73 & 6.37 & 5.60 & 1.32 & 1.39 \\ 
            67 & 11.28 & 8.72 & 0.78 & 0.86 & 1.64 & 1.39 & 0.61 & 0.68 & 6.91 & 5.93 & 1.28 & 1.38 \\ 
            68 & 14.79 & 10.35 & 0.75 & 0.85 & 1.72 & 1.42 & 0.52 & 0.62 & 7.65 & 6.37 & 1.24 & 1.35 \\ 
            69 & 21.60 & 14.62 & 0.72 & 0.84 & 1.82 & 1.46 & 0.40 & 0.49 & 8.65 & 7.22 & 1.17 & 1.31 \\ 
            70 & 73.31 & 109.98 & 0.60 & 0.75 & 2.18 & 1.73 & 0.16 & 0.10 & 11.39 & 10.77 & 0.93 & 1.05 \\ 
            71 & 178.23 & 559.83 & 0.51 & 0.63 & 2.45 & 2.09 & 0.07 & 0.02 & 13.01 & 13.65 & 0.76 & 0.70 \\ 
            72 & 8.67 & 7.57 & 0.80 & 0.87 & 1.57 & 1.37 & 0.71 & 0.73 & 6.15 & 5.56 & 1.33 & 1.40 \\ 
            73 & 9.09 & 7.82 & 0.80 & 0.87 & 1.59 & 1.38 & 0.69 & 0.72 & 6.29 & 5.64 & 1.32 & 1.39 \\ 
            74 & 9.66 & 8.16 & 0.79 & 0.86 & 1.60 & 1.38 & 0.67 & 0.71 & 6.46 & 5.76 & 1.31 & 1.39 \\ 
            75 & 10.38 & 8.69 & 0.79 & 0.86 & 1.62 & 1.39 & 0.64 & 0.68 & 6.67 & 5.92 & 1.30 & 1.38 \\ 
            76 & 12.32 & 10.37 & 0.77 & 0.85 & 1.67 & 1.42 & 0.58 & 0.61 & 7.15 & 6.37 & 1.27 & 1.35 \\ 
            77 & 13.70 & 11.73 & 0.76 & 0.85 & 1.69 & 1.43 & 0.54 & 0.57 & 7.43 & 6.68 & 1.25 & 1.34 \\ 
            78 & 20.82 & 22.63 & 0.73 & 0.82 & 1.80 & 1.52 & 0.41 & 0.36 & 8.52 & 8.16 & 1.18 & 1.26 \\ \hline 
            79 & 8.68 & 6.22 & 0.86 & 0.91 & 1.39 & 1.25 & 0.68 & 0.76 & 5.93 & 4.71 & 1.38 & 1.50 \\ 
            80 & 13.16 & 8.55 & 0.84 & 0.90 & 1.46 & 1.27 & 0.53 & 0.63 & 7.00 & 5.42 & 1.32 & 1.48 \\ 
            81 & 32.18 & 19.66 & 0.79 & 0.88 & 1.62 & 1.33 & 0.28 & 0.36 & 9.16 & 7.05 & 1.16 & 1.41 \\ 
            82 & 292.08 & 369.64 & 0.66 & 0.80 & 2.00 & 1.58 & 0.04 & 0.03 & 12.91 & 10.63 & 0.79 & 1.11 \\ \hline 
            86 & $-$ & 8.52 & $-$ & 0.85 & $-$ & 1.43 & $-$ & 0.71 & $-$ & 6.07 & $-$ & 1.34 \\ 
            87 & $-$ & 13.69 & $-$ & 0.81 & $-$ & 1.55 & $-$ & 0.55 & $-$ & 7.53 & $-$ & 1.23 \\ 
            88 & $-$ & 17.43 & $-$ & 0.77 & $-$ & 1.66 & $-$ & 0.48 & $-$ & 8.38 & $-$ & 1.16 \\ 
            89 & $-$ & 8.32 & $-$ & 0.85 & $-$ & 1.42 & $-$ & 0.72 & $-$ & 5.98 & $-$ & 1.35 \\ 
            90 & $-$ & 8.99 & $-$ & 0.85 & $-$ & 1.43 & $-$ & 0.30 & $-$ & 6.19 & $-$ & 1.34 \\ 
            91 & 31.31 & 10.97 & 0.62 & 0.88 & 2.12 & 1.32 & 0.34 & 0.60 & 10.58 & 6.58 & 0.94 & 1.33 \\ 
            92 & 207.85 & 21.53 & 0.26 & 0.86 & 3.23 & 1.39 & 0.08 & 0.39 & 16.25 & 8.40 & 0.31 & 1.21 \\ 
            95 & 12.35 & 7.60 & 0.75 & 0.90 & 1.73 & 1.28 & 0.61 & 0.74 & 7.58 & 5.60 & 1.19 & 1.39 \\ 
            96 & 17.46 & 9.41 & 0.71 & 0.89 & 1.84 & 1.30 & 0.49 & 0.66 & 8.64 & 6.19 & 1.11 & 1.35 \\ 
            97 & 92.78 & 42.46 & 0.51 & 0.85 & 2.47 & 1.43 & 0.14 & 0.23 & 13.16 & 9.57 & 0.70 & 1.14 \\ 
            98 & 269.15 & 467.42 & 0.37 & 0.76 & 2.90 & 1.71 & 0.06 & 0.03 & 15.30 & 13.64 & 0.45 & 0.70 \\ \hline 
            101 & 457.38 & 26.18 & 0.67 & 0.76 & 1.97 & 1.70 & 0.03 & 0.24 & 12.77 & 6.20 & 0.82 & 1.60 \\ \hline 
            102 & 55.36 & 5.21 & 0.60 & 0.85 & 2.18 & 1.41 & 0.20 & 0.76 & 11.10 & 3.94 & 0.95 & 1.61 \\ \hline 
            103 & 26.76 & 6.44 & 0.86 & 0.84 & 1.40 & 1.45 & 0.33 & 0.67 & 8.79 & 4.32 & 1.19 & 1.61 \\ 
            104 & $-$ & 1113.80 & $-$ & 0.61 & $-$ & 2.15 & $-$ & 0.01 & $-$ & 7.31 & $-$ & 1.58 \\ \hline 
            106 & 16.63 & 3.34 & 0.71 & 0.88 & 1.84 & 1.33 & 0.51 & 0.94 & 8.52 & 3.15 & 1.12 & 1.61 \\ 
            \hline
            \multicolumn{13}{l}{
            \parbox{\linewidth}{\vspace{5pt}\textit{Notes:} 
Column 1 gives the case number for cross-referencing with Table \ref{table:2_3_physical_table}. Columns 2 and 3 give the global, mass-weighted N/C ratios, relative to solar, for final bound components 1 and 2, respectively. Columns 3 and 4, columns 5 and 6, columns 7 and 8, columns 9 and 10, and columns 11 and 12 are the same for hydrogen, helium, carbon, nitrogen, and oxygen, respectively.}
            \vspace{5pt}}\\
        \enddata
        \vspace{0.2cm}
    \end{deluxetable*}

As a self-consistency check, we confirm in post-processing that the brief increase in the CNO-cycle energy production during the Case 32 and 82 collisions is negligible compared to the energy generated by the CNO cycle over the lifetime of the parent star. In particular, by following the density and temperature evolution of the particles during the collision, we estimate the amount of compositional change from the collision to be comparable to what would occur over a few hundred to a few thousand years in the star's evolution, allowing the assumption of constant composition during the hydrodynamic simulation to remain valid.

\section{Implications of Collisionally Formed Stripped Stars}\label{sec:implications}

\subsection{Total Tidal Disruption Events}\label{subsec:tdes}

    \begin{figure*}
        \centering
        \includegraphics[width=\linewidth]{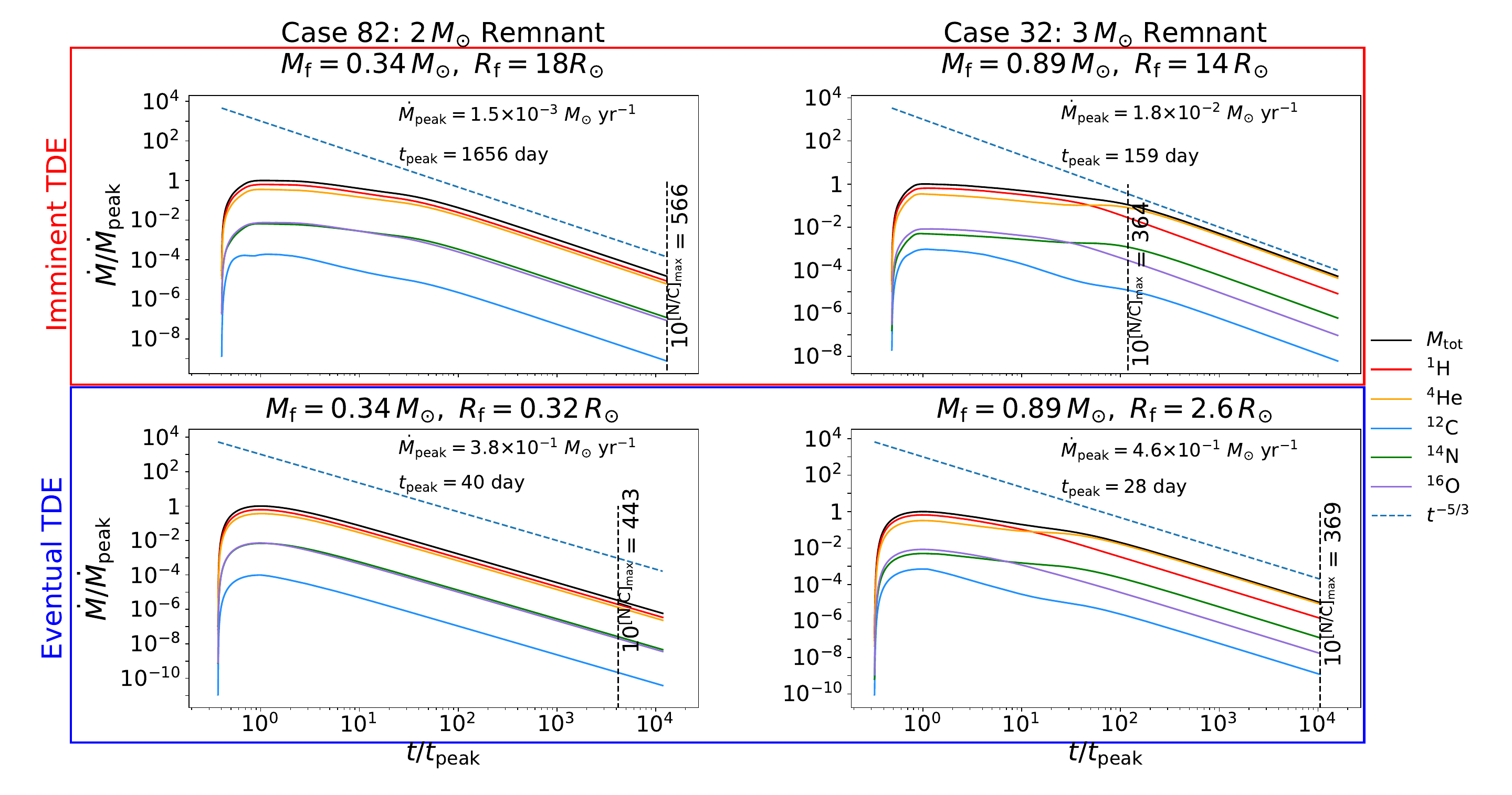}

        \caption{An estimate of the fallback material in the complete TDE of collisionally stripped stars. We consider the product of the $2\,M_{\odot}$ star from Case 82 \textbf{(left)} and the product of the $3\,M_{\odot}$ star from Case 32 \textbf{(right)}, completely disrupted by a supermassive black hole with mass $4.3\times10^6\,M_{\odot}$.
        The ``imminent TDE" fallback rates (top row) are for models that are in hydrostatic equilibrium but still thermally bloated from the collision, while the ``eventual TDE" fallback rates (bottom row) are for the same models once they have been relaxed into {\tt MESA}, reaching their smallest radius. The $0.34\,M_{\odot}$ and $0.89\,M_{\odot}$ stars took $\sim10^7\,{\rm yr}$ and $\sim10^5$ years, respectively, to reach this point. All of these scenarios result in [N/C]$_{\rm max}>$[N/C]$_{\rm ASASSN-14li}$ for the constraints set in both \citet{Yang17} (31.6) and \citet{Miller+23} (251).
        The star from Case 32 is especially interesting because it has a mass-weighted average [N/C]$<$[N/C]$_{\rm ASASSN-14li}$.
        The analytic prescription enforces that exactly half of the star, from the outermost point to the center, falls back to the black hole. The maximum N/C value is taken from a time after the peak.}
        \label{fig:tde_fallback}
    \end{figure*}

The stars created from these high-velocity collisions have atypical chemical abundances compared to non-stripped stars of similar mass: they have much less hydrogen- and carbon-rich material and much more helium- and nitrogen-rich material.
The tidal disruption of these stars could therefore make them ideal candidates for progenitors of TDEs with anomalous chemical abundances, such as ASASSN-14li, PTF15af, and iPTF16fnl \citep{Cenko+16,Kochenek+16,Blagorodnova+17,Blagorodonva+19,Yang17}.

We present an analytical estimate of the time evolution of the composition of the fallback material of TDEs of the collision products following \citet{Gallegos-Garcia+18}.
Specifically, the shape of our fallback curves is given by the approach of \citet{Gallegos-Garcia+18}, scaled to the estimate of the peak fallback time $t_{\rm peak}$ introduced by \citet{2022MNRAS.517L..26C} and further analyzed by \citet{2024ApJ...961L...2B}. The maximum fallback rate $\dot M_{\rm peak}$ is set such that the total mass falling back to the SMBH is half the star mass.
Although a full hydrodynamical simulation is required to track the details of TDEs, such as mixing of material, the analytical estimate should broadly capture the general features. Compared to hydrodynamical simulations, the analytical estimate shows good agreement at late times \citep{Law-Smith+19}.

We begin by sphericalizing the data, creating a 1-D HSE model and a MS {\tt MESA} model of the data as outlined in Section \ref{subsec:sph_to_mesa}.
The consistency between the raw SPH particle data, the smoothed curve SPH inputs and the resulting {\tt MESA} evolution used to generate the MS model is discussed in Appendix \ref{app:sphericalization_details}.

Figure \ref{fig:tde_fallback} shows the fallback debris of the remnant of the $2\,M_{\odot}$ star from Case 82 with mass-weighted average $10^{\rm[N/C]}=370$ and the remnant of the $3\,M_{\odot}$ star from Case 32 with $10^{\rm[N/C]}=167$.
We assume complete disruption, which requires that the penetration factor $\beta$, defined as the ratio of the tidal radius to the distance of closest approach $r_{\rm T}/r_{\rm p}$, exceed a critical value $\beta_{\rm c}$.
From \citet{2022MNRAS.517L..26C,Ryu+20,Li-Xin+02},
$\beta_{\rm c} \approx [\rho_{\rm c}/(4 \rho_*)]^{1/3}$ for a star of central density $\rho_{\rm c}$ and average density $\rho_*$.
For the Case 82 product, $\beta_{\rm c}\approx 7$ for the HSE model and 1.3 for the MS model.  The corresponding $\beta_{\rm c}$ values for the Case 32 product are approximately 14 and 9.
Because our approximate treatment assumes the star is completely disrupted at the tidal radius, it (for a given star-black hole combination) gives the same results regardless of $\beta$. Such behavior is in qualitative agreement with the numerical results of \citet{GuillochonRR13}, who show that the fallback of the debris in a complete TDE is relatively insensitive to the penetration factor for $\beta$.

We show that both the timescale and the shape of the fallback of the debris are sensitive to the density profile of the star; a denser star will experience a steeper fallback curve with shorter timescales, while a puffy, low-density object, like collision products shortly after their formation, will experience a shallower fallback curve with much longer timescales.
Following our SPH collision calculations, the surviving stars have been forced into hydrostatic equilibrium.
However, these stars will only relax back to thermal equilibrium on a longer timescale, decreasing in size and increasing in mean density as a result.

If the final orbital trajectory of the star will lead the newly stripped star towards the SMBH in less time than the Kelvin-Helmholtz timescale, then the top curves may be more consistent with the fallback.
However, if the stripped star relaxes into thermal equilibrium before it undergoes a TDE, then the bottom curves may be more consistent with the fallback.

Regardless of the shape of the fallback curve, the composition of the fallback material consistently has more nitrogen than carbon, and at late times, the N/C abundance ratio is consistent with the constraints described in \cite{Miller+23}.

Although the [N/C] abundance ratio of several products from this work are consistent with the constraints set forth in both \cite{Yang17} and \cite{Miller+23}, the nitrogen abundance for the stars we are analyzing are still lower than the fitted constraints of nitrogen proposed in \cite{Miller+23}.
However, uncertainties in the individual elemental abundances are much higher than those in the ratio of the abundances \citep[especially given the uncertainty in the gas conditions and therefore ionization states around the SMBH, e.g.][]{batra_metallicities_2014, Yang17}, meaning that the stripped stars created in this work may still be reasonable progenitors of the stripped star TDEs.
\subsection{Partial Tidal Disruption Events}\label{subsec:ptdes}
Observations of repeating partial TDEs (pTDEs) with similar amplitude flares provide an additional motivation for stellar collisions in galactic nuclei.
One such example is ASASSN-14ko, a candidate repeating partial disruption that has been observed to produce very consistent optical and UV flares every 114 days since it was discovered in 2014 \citep{Payne+21}.
This transient is consistent with the disruption of an extended donor star of mass $\gtrsim1\,M_{\odot}$, radius $\approx\,10R_{\odot}$, and density concentration ratio $\rho_{\rm c}/\bar\rho > 10^3$
\citep{Liu+23}
orbiting an SMBH of mass $\sim 10^{7.86}\,M_{\odot}$ \citep{Payne+21}.
An extended star is ideal for repeated pTDEs because, unlike MS stars which experience significant structural changes from the encounter, puffy stars may only experience mass loss from their
low-density outer layers, preserving the internal structures \citep{Law-Smith+20}.
This semi-constant structure over many pTDEs allows the magnitude of each flare to vary much less than repeated pTDEs of
stars more homogeneous in density.
Here we provide a simple analysis of the ASASSN-14ko orbit and compare the scenario to our simulation results.
With an orbital period of 114 days \citep{Payne+21}, we use Kepler’s Third Law to find a semimajor axis $a \sim200$\,AU.
To achieve small but non-zero mass transfer to the SMBH, we assume that the orbit has a pericenter distance $r_{\rm p}\sim 2\,r_{\rm T}\sim 40$\,AU as defined by the stellar constraints from \citet{Liu+23}.
These values of $r_{\rm p}$ and $a$ correspond to an eccentricity $\sim0.8$ and an apocenter distance $r_{\rm a}\sim 340$\,AU.

The donor star in ASASSN-14ko could be extended simply because it is a normal evolved star such as a giant.  Another possibility, which we consider here, is that it has been puffed up by a stellar collision.  Such a collision could have occurred at an orbital separation comparable to the current orbital separation of ASASSN-14ko, where typical $v_\infty$ values exceed $\sim10^4$\,km\,s$^{-1}$.
Several of the high-velocity collisions ($v_\infty\geq 10^4$\,km\,s$^{-1}$) that we present (Cases 45, 46, 49, 50, and 78) leave behind at least one extended star of mass $\gtrsim 1\,M_\odot$. As the impact in these cases is off-axis enough to avoid complete disruption, the post-collision velocities of the resulting stars are not substantially different from that of the parent stars from which they came: the collision typically decreases the speed by less than a few percent in these scenarios.  Such collisions, therefore, could allow a star that is already closely orbiting the SMBH to remain on essentially the same bound orbit, but extend the stellar radius enough to initiate repeated pTDEs \citep[e.g.,][]{Ryu_2020,Kiroglu_2023}.

Collision products such as that of the $3\,M_{\odot}$ TAMS star, although puffy, retain high central densities, allowing them to survive repeated pTDEs \citep{2024arXiv240601670L}. 
For example, even after being brought to hydrostatic equilibrium, the Case 46 collision product has a radius of $16\,R_{\odot}$, roughly consistent with the radius the star estimated by \citet{Liu+23} for ASASSN-14ko, and a density concentration ratio $\rho_{\rm c}/\bar\rho\sim 10^4$, consistent with the  lower limit of $10^3$ placed on the donor star in ASASSN-14ko by \citet{Liu+23}.
Additionally, frequent pericenter passages slow down the thermal relaxation of the star, allowing it to remain bloated and experience repeated pTDEs.
Due to the high frequency of collisions close to central SMBHs, such collision products may be able to explain cases like ASASSN-14ko.
Further hydrodynamic modeling of the orbit of inflated stellar collision products is necessary to validate the possibility of repeated pTDE flares.

\subsection{G Objects}\label{subsec:g-objects}

Stellar collisions may produce other objects of interest in galactic nuclei. Specifically, peculiar dust and gas-enshrouded stellar objects, called G~objects, have been observed in the Milky Way's nuclear star cluster: currently, six G~objects are known \citep{Ciurlo+20}. Their origin remains unknown, though several formation channels have been proposed in the literature \citep[e.g.,][]{Burkert+12,Schartmann+12,Zajacek+17,Madigan+17,OwenLin23}. The proposed formation channels include secular binary mergers \citep[e.g.,][]{Witzel+14,Prodan+15,Stephan+16,Stephan+19,2020ApJ...901...44W} and stellar mergers from direct collisions of single stars \citep{Rose+23}. Here we evaluate the potential of high-speed collisions, often producing stripped stars as demonstrated in this paper, to contribute to the G~object population.
Immediately post-collision, these stars are puffy and extended, similar to G~objects, as they have not yet contracted toward thermal equilibrium.

Following the calculation of \citet{Rose+23}, we estimate that collisions could explain some fraction of the G objects in the Galactic Center depending on the slope of the stellar density profile, $\gamma$.
The number of G~objects can be calculated as $N_{\rm G} = r_{\rm G} t_{\rm puffy}$, where $r_{\rm G}$ is the rate of G~object formation and $t_{\rm puffy}$ is the average lifetime of a G~object, that is, the time that it remains in a distended form.  The rate is calculated as
\begin{align}
r_{\rm G} = \int_{r_{\rm min}}^{r_{\rm max}}k n \frac{1}{t_{\rm coll}}4 \pi r^2 dr,
\end{align}
where $r_{\rm min}$ and $r_{\rm max}$ are the minimum and maximum radii, respectively, at which we are considering collisions, 
where the collision timescale $t_{\rm coll}$ is given by Equation~(\ref{equation:tcoll}), and
where $k=k(r)$ represents the average number of G~objects formed per collision at radius $r$.  The function $k(r)$ is bounded by 0 (corresponding to the complete disruption of both stars) and 2 (corresponding to both stars becoming a G~object).  For simplicity, we set $k=1$, corresponding to an average of one G~object being made per collision.

When evaluating the collision timescale $t_{\rm coll}$ and the number density $n(r)$, we assume a stellar population comprised of solar mass and solar radii stars.
With the density profile $\rho(r)$ from \citet{Genzel+10}, the number density profile is therefore $n=\rho/M_\odot=1.35\times10^6(r/0.25\,{\rm{ pc}})^{-\gamma}\,\rm{pc}^{-3}$ for $r\lesssim 0.25$\,pc.
Assuming $r_{\rm max} >> r_{\rm min}$, we integrate to find
that, if $\gamma=1.25$,
\begin{equation}
    N_{\rm G}=\frac{t_{\rm puffy}}{10^4\,{\rm yr}}\left[0.07 \frac{\log\left(\frac{r_{\rm max}}{r_{\rm min}}\right)}{\log\left(\frac{0.25\,{\rm pc}}    {10^{-4}\,{\rm pc}}\right)} + 0.06 \frac{r_{\rm max}}{0.25\,{\rm pc}}\right],
\label{equation:NG1}
\end{equation}
if $\gamma=1.5$,
\begin{equation}
N_{\rm G}=\frac{t_{\rm puffy}}{10^4\,{\rm yr}}\left[0.9\left(\frac{r_{\rm min}}{10^{-4}\,{\rm pc}}\right)^{-1/2}+0.1\left(\frac{r_{\rm max}}{0.25\,{\rm pc}}\right)^{1/2}\right],
\label{equation:NG2}
\end{equation}
if $\gamma=1.75$,
\begin{equation}
    N_{\rm G}=\frac{t_{\rm puffy}}{10^4\,{\rm yr}}\left[20 \left(\frac{r_{\mathrm{min}}}{10^{-4}\,\mathrm{pc}}\right)^{-1} + 0.5 \frac{\log\left(\frac{r_{\mathrm{max}}}{r_{\mathrm{min}}}\right)}{\log\left(\frac{0.25\,\mathrm{pc}}{10^{-4}\,\mathrm{pc}}\right)}\right],
\label{equation:NG3}
\end{equation}
and if $1.25<\gamma<1.75$,
\begin{equation}
\begin{array}{ll}
    N_{\rm G}=\frac{t_{\rm puffy}}{10^4\,{\rm yr}}\left[0.014 \frac{(1+\gamma)^{-1/2}}{2\gamma-2.5} \left(\frac{r_{\mathrm{min}}}{0.25\,\mathrm{pc}}\right)^{2.5-2\gamma}\right. \\
    \hspace{1em}+
    \left.0.037 \frac{(1+\gamma)^{1/2}}{3.5-2\gamma} \left(\frac{r_{\mathrm{max}}}{0.25\,\mathrm{pc}}\right)^{3.5-2\gamma}\right].
\label{equation:NG4}
\end{array}
\end{equation}
The sum of terms on the right-hand side arises from the sum in Equation~(\ref{equation:tcoll}), with the second term accounting for gravitational focusing.

The estimated number of G~objects formed through this collisional channel depends sensitively on the parameters $\gamma$, $r_{\rm min}$, and $t_{\rm puffy}$, all of which are highly uncertain.  Observationally, it is difficult to know how closely to Sgr A* stars exist, but clearly, the value $r_{\rm min}$ cannot be less than the tidal disruption radius, which is on the order of $10^{-5}$\,pc for Sun-like stars.
The parameter $t_{\rm puffy}$ is likely roughly the thermal time of the collision product, which may be of order $10^4$\,yr based on the thermal timescale in the outer layers of collision products formed in parabolic, head-on collisions \citep{1997ApJ...487..290S}.
For $r_{\rm min}=10^{-4}$\,pc,
$r_{\rm max}=0.25$\,pc, and $t_{\rm puffy}=10^4$\,yr, Equations~(\ref{equation:NG1}), (\ref{equation:NG2}), and (\ref{equation:NG3}) predict 0.1, 1, or 20 G~objects.
If in fact the true relaxation time $t_{\rm puffy}$ is ten times larger, then we predict as many as 1, 10, or 200 objects for the same $\gamma$ values. We plan to investigate the details more rigorously in a future dedicated study.

\section{Discussion} \label{sec:discussion}

Our simulations have shown that high-velocity stellar collisions can create stripped stars.
Our collision simulations result in four types of qualitative outcomes:
\begin{enumerate}
    \item \textbf{Two Stripped Stars:} For collisions between two intermediate-mass stars, two products may survive with high mass loss, both with an elevated N/C abundance ratio.
    \item \textbf{One Stripped Star:} One product remains with an elevated N/C abundance ratio. The other star in the collision may have been completely disrupted or it may not have had enough mass stripped from it to have an elevated N/C abundance ratio.
    \item \textbf{No Stripped Stars:} One or two stars remain, but neither of them have experienced significant mass loss and an elevated N/C abundance ratio. This is often the case for grazing or lower speed collisions, which do not provide enough energy to strip substantial mass. This case also includes merger products.
    \item \textbf{Complete Disruption:} Both stars in the collision are completely destroyed. This typically happens for higher velocity or more direct collisions.
\end{enumerate}

We find that there is a region in the $r_{\rm p}$ vs $v_{\infty}$ parameter space in which the first two outcomes occur.
Velocities greater than $\sim2000\rm\,km\,s^{-1}$ are able to eject enough mass from the stars to significantly increase the average N/C abundance ratio to be consistent with \citet{Yang17}.
These velocities also show agreement with mass loss from high velocity stellar collisions presented in \citet{FreitagBenz2005}.
Collisions with velocites of this scale are expected to occur within the inner $\sim10^{-2}$\,pc of the Galactic Center, falling within the orbital constraints of S-stars like S0-2.
Furthermore, these velocites may be present at greater distances for more massive galaxies.
The formation of these stripped stars may be sensitive to the age, mass, and collision history of the star involved, requiring investigation of modeling of a broader parameter space.

We present an analytical estimate of the compositional fallback of several stripped stars of interest.
Although the shape and timescale of the fallback of the stellar debris is sensitive to the internal structure of the star, the estimates that we present all show an N/C abundance ratio of the fallback material that is consistent with the constraints set forth by \citet{Yang17,Miller+23}.
The maximum N/C abundance ratios presented in the fallback are greater than the globally averaged N/C for the entire star; especially at late times, there is less dilution from the outer, hydrogen- and carbon-rich envelope. This implies that stars that do not experience enough mass loss to significantly elevate the global N/C abundance ratio may still be candidates for stripped star TDEs.

A greater understanding of the relationship between stellar dynamics, stellar evolution, and hydrodynamics is necessary to better inform the interactions.
For example, thermally bloated collision products in tight orbits around an SMBH may be able to explain pTDEs like ASASSN-14ko. They may also be able to explain G objects in the Galactic Center.
Additionally, simplifications made to the stellar profiles to predict the TDE fallback remove information regarding anisotropies in the high N/C material throughout the collision product.
Direct hydrodynamical modeling of TDEs involving collision products may remove these simplifications and provide further insight into the nature of collisionally-formed stripped star TDEs.

Most of the high N/C stars presented in this work originate from intermediate mass stars due to their nitrogen enhancement from the CNO cycle.
However, solar and low mass stars may be important in the formation of collisionally-formed stripped stars for two reasons.
First, collisions between solar-type and intermediate mass stars were shown to be energetic enough to strip substantial mass from the higher mass star.
However, even lower mass stars may also be able to strip significant mass from the intermediate mass stars that are presented.
This could be explored with further hydrodynamical modeling.
Second, we show that solar-type stars may also yield a low-mass remnant with an elevated N/C abundance ratio.
Although it does not experience significant CNO processing, carbon depletion in the core of the $1\,M_{\odot}$ star is more substantial than higher-mass stars, allowing the N/C abundance ratio to remain large.
Interactions strictly with solar and low mass stars may yield fewer complete disruption cases and could be another, more plausible formation channel of high N/C objects in galactic nuclei, especially due to the higher likelihood of the existence of lower mass stars.
However, further hydrodynamical modeling of these collisions is necessary to better understand the structure and composition of these collision products.

Summary data from this work, {\tt MESA} inlists, the source code for this version of {\tt StarSmasher}, and the post-processing codes are available on Zenodo under a GNU General Public License: \dataset[doi:10.5281/zenodo.13883137]{https://doi.org/10.5281/zenodo.13883137}.

\begin{acknowledgments}
We thank Giacomo Fragione, Ilya Mandel, Jon Miller, Carl Rodriguez, Taeho Ryu, and Meng Sun for helpful conversations, and the anonymous referee for valuable feedback.
This work was initiated while several of the authors (SCR, BM, KK, ER-R, FAR) were at the Aspen Center for Physics, which is supported by NSF grant PHY-2210452.
MG-G is grateful for the support from Northwestern University's Presidential Fellowship.
CFAG was supported as a summer research undergraduate at Northwestern University through NASA grant 80NSSC20M0046 awarded to the Illinois/NASA Space Grant Consortium.
Support for KK was provided by NASA Hubble Fellowship grant HST-HF2-51510 awarded by the Space Telescope Science Institute, which is operated by the Association of Universities for Research in Astronomy, Inc., for NASA, under contract NAS5-26555.
BM is grateful for support from the Carnegie Theoretical Astrophysics Center.
ER-R acknowledges support from the Heising-Simons Foundation, NSF grants AST-2150255 and AST-2307710, Swift grants 80NSSC21K1409 and 80NSSC19K1391, and Chandra grant 22-0142.
FAR acknowledges support from NASA grants 80NSSC21K1722 and 80NSSC22K0722, and NSF grant AST-2108624.
SCR acknowledges support from the CIERA Lindheimer Fellowship.
This research was supported in part through the computational resources and staff contributions provided for the Quest high-performance computing facility at Northwestern University, which is jointly supported by the Office of the Provost, the Office for Research, and Northwestern University Information Technology.
Some of our computations were also conducted at the Resnick High Performance Computing Center, a facility supported by the Resnick Sustainability Institute at the California Institute of Technology.
This work made use of
NumPy \citep{harris2020array}, Matplotlib \citep{Hunter:2007}, SciPy \citep{2020SciPy-NMeth}, Pandas \citep{reback2020pandas}, Astropy \citep{astropy:2013, astropy:2018, astropy:2022}, and the SPLASH visualization software \citep{2007PASA...24..159P}.
\end{acknowledgments}

\appendix
\section{Stretchy HCP Lattice}\label{app:stretchy_hcp}

    \begin{figure}\label{fig:stretchyhcp}
        \centering
        \includegraphics[width=0.8\linewidth]{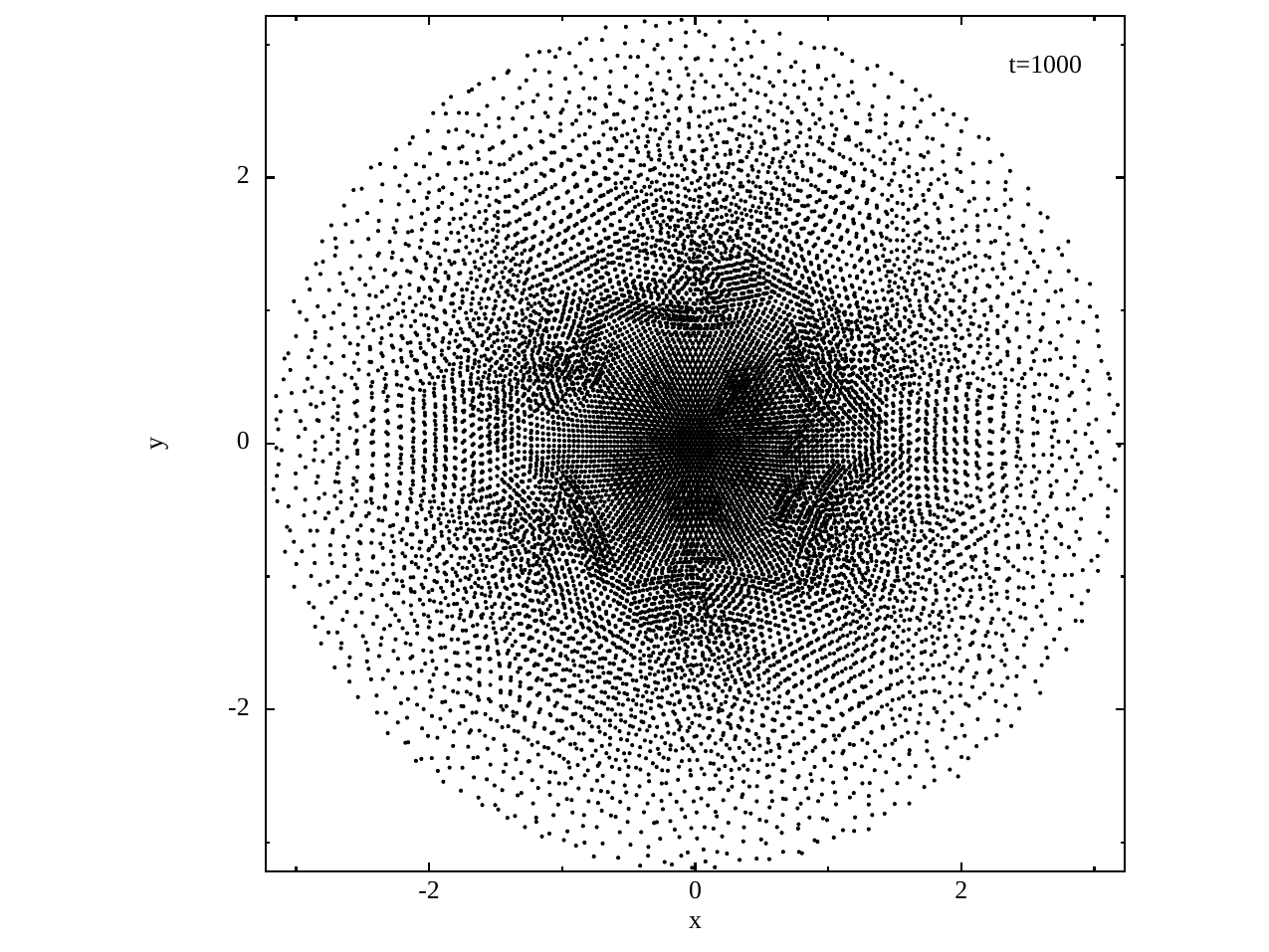}
        \caption{A cross-sectional slice, covering $|z|<0.1 R_\odot$ (where $z=0$ is the midplane), of the relaxed $3\,M_{\odot}$ star. A higher resolution in the center allows for a more accurate internal structure to be constructed following the collision.
        }
    \end{figure}

A common way to initialize an SPH model of a star is with a constant number density of particles. Certain lattice configurations for these particles, such as hexagonal close-packed (hcp), can be stable against perturbations \citep{1999JCoPh.152..687L}, making them a natural choice.  Using a constant number density of particles allows for uniform spatial resolution throughout the mass distribution. This approach therefore has the advantage of computational efficiency, at least initially, as no small region of the simulation will be required to take particularly small timesteps and slow down the entire computation. Constant number density particles however have the disadvantage of giving poor mass resolution in high-density regions. In stars that have significant density gradients, such as the TAMS stars considered in this paper, very few particles would be used to represent a large percentage of the overall mass.

Another option for initializing SPH models is to choose a uniform particle mass. In such a scenario, the number density of SPH particles is proportional to the local mass density. This approach focuses computational resources equitably by automatically enforcing uniform mass resolution. The disadvantage of such an approach is that particles can be placed extremely close together in high-density regions causing the timestep to be exceedingly small.

In practice, a compromise between these two approaches is often the best option in stars with a large density contrast. Instead of having the number density be constant or proportional to the mass density, we can take the particle number density $n_{\rm part}$ to be proportional to the mass density $\rho$ raised to a power, as in \cite{FreitagBenz2005}:
\begin{equation}
n_{\rm part} \propto \rho^\alpha. \label{numberDensity}
\end{equation}
The free parameter $\alpha$ allows for treatments that smoothly transition from the constant number density case ($\alpha=0$) to the equal particle case ($\alpha=1$).  In the input file of {\tt StarSmasher}, the parameter $\alpha$ is called {\tt equalmass}, and for the parent star models in this paper, we have chosen $\alpha=0.4$.  The desired density profile $\rho = m_{\rm part} n_{\rm part}$ is still achieved, as particle masses are chosen such that $m_{\rm part} \propto \rho^{1-\alpha}$.

One straightforward way to realize equation (\ref{numberDensity}) is with a Monte Carlo approach. However, this results in stochastic fluctuations to the mass distribution and therefore to Poisson shot noise. The resulting motion of SPH particles can be substantial and, even if damped, can lead to a spurious restructuring of the stellar interior.

In this paper, we introduce a new alternative means for achieving equation \ref {numberDensity}, which we dub the stretchy HCP method. The basic idea is that we adjust the lattice cell spacing throughout the star in such a way that Equation (\ref{numberDensity}) is achieved.  To accomplish this, we first note that the desired number of particles $N$ must satisfy
\begin{equation}
N = \int_0^{R_{\rm edge}} n_{\rm part} 4 \pi r^2 dr = n_{\rm c} \int_0^{R_{\rm edge}}  \left(\frac{\rho(r)}{\rho(0)}\right)^\alpha 4 \pi r^2 dr,  \label{nc}
\end{equation}
where $n_{\rm c}$ is the number density at the center of the star (the origin) and $R_{\rm edge}$ is the radius out to which particles will be placed (about two smoothing lengths less than the full radius of the star). Equation (\ref{nc}) allows us to solve for the central number density $n_{\rm c}$, as the stellar model sets the density profile $\rho(r)$ and the user chooses $N$ and $\alpha$.

The particle positions in the stretchy hcp lattice are mapped from positions in a hypothetical unstretched hcp lattice of uniform number density $n_{\rm c}$ (corresponding to a cell volume of $2/n_{\rm c}$, as there are two particles per primitive unit cell in an hcp lattice).  This mapping is based on there being the same number of particles in a spherical shell of radius $r_i$ and thickness $dr_i$ in the unstretched lattice as there is in a shell of radius $r_f$ and thickness $dr_f$ in the stretched lattice:
$n_{\rm c} 4 \pi r_{\rm i}^2 dr_{\rm i} = n_{\rm part} 4 \pi r_{\rm f}^2 dr_{\rm f}.$
Requiring that the central number density of the unstretched and stretched lattices be equal yields
\begin{equation}
r_{\rm i}^2 \frac{dr_{\rm i}}{dr_{\rm f}} = \left(\frac{\rho(r_{\rm f})}{\rho(0)}\right)^\alpha r_{\rm f}^2. \label{difeq}
\end{equation}
This differential equation is solved to give $r_{\rm i} = r_{\rm i}(r_{\rm f})$, subject to the initial condition $r_{\rm i}(0)=0$, and then that result is inverted to give $r_{\rm f} = r_{\rm f}(r_{\rm i})$.
The factor $r_{\rm f}/r_{\rm i}$ provides the mapping from the unstretched to the stretched lattice.
In particular, a lattice site at position $\vec{r}_{\rm i}$
in the unstretched lattice corresponds to the position $r_{\rm f}\vec{r}_{\rm i}/r_{\rm i}$
in the stretched lattice, with $r_{\rm i} = |\vec{r}_{\rm i}|$.

\section{Further Details on Sphericalization}\label{app:sphericalization_details}

\begin{figure}
    \includegraphics[width=0.495\linewidth]{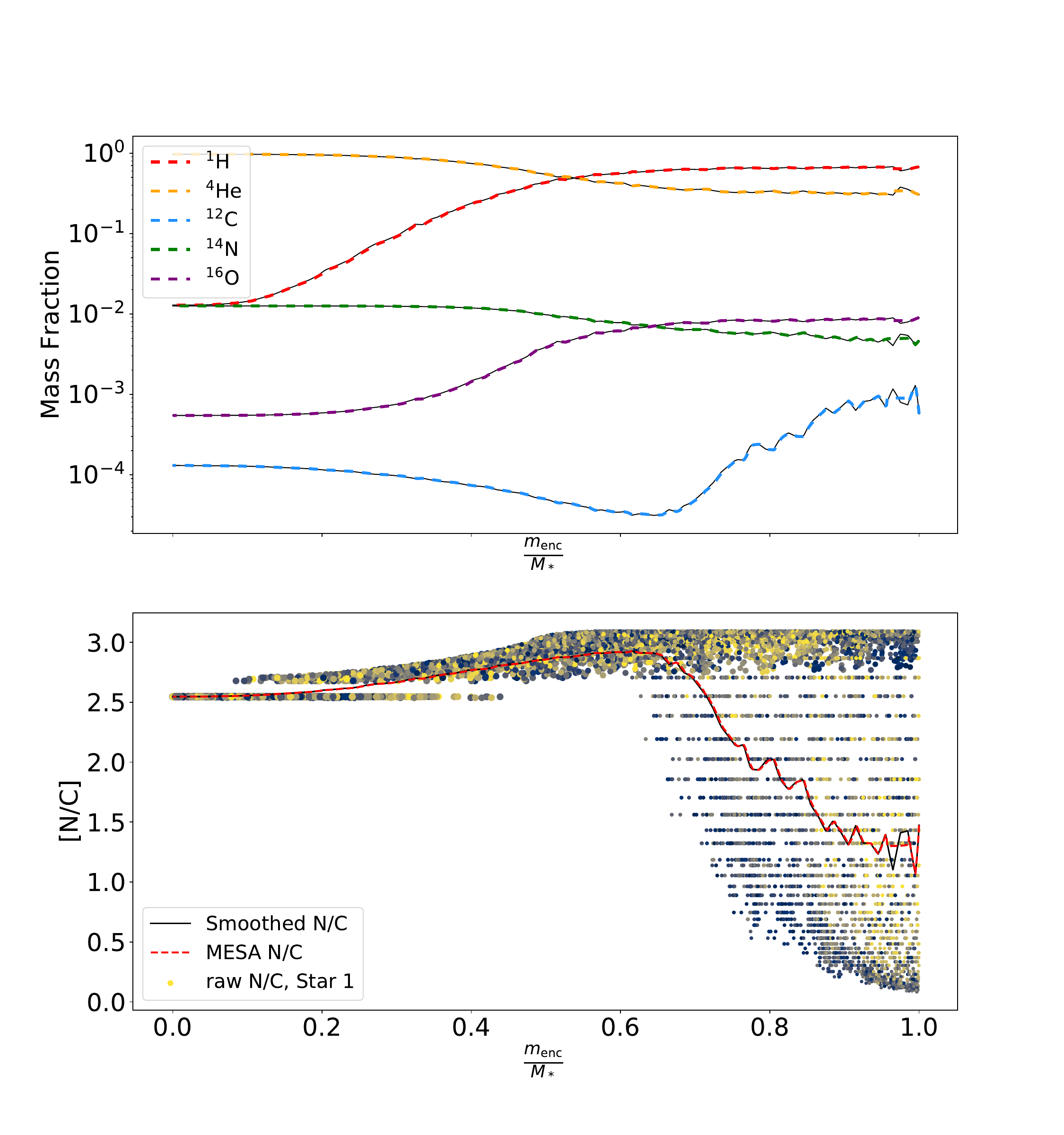}
    \includegraphics[width=0.495\linewidth]{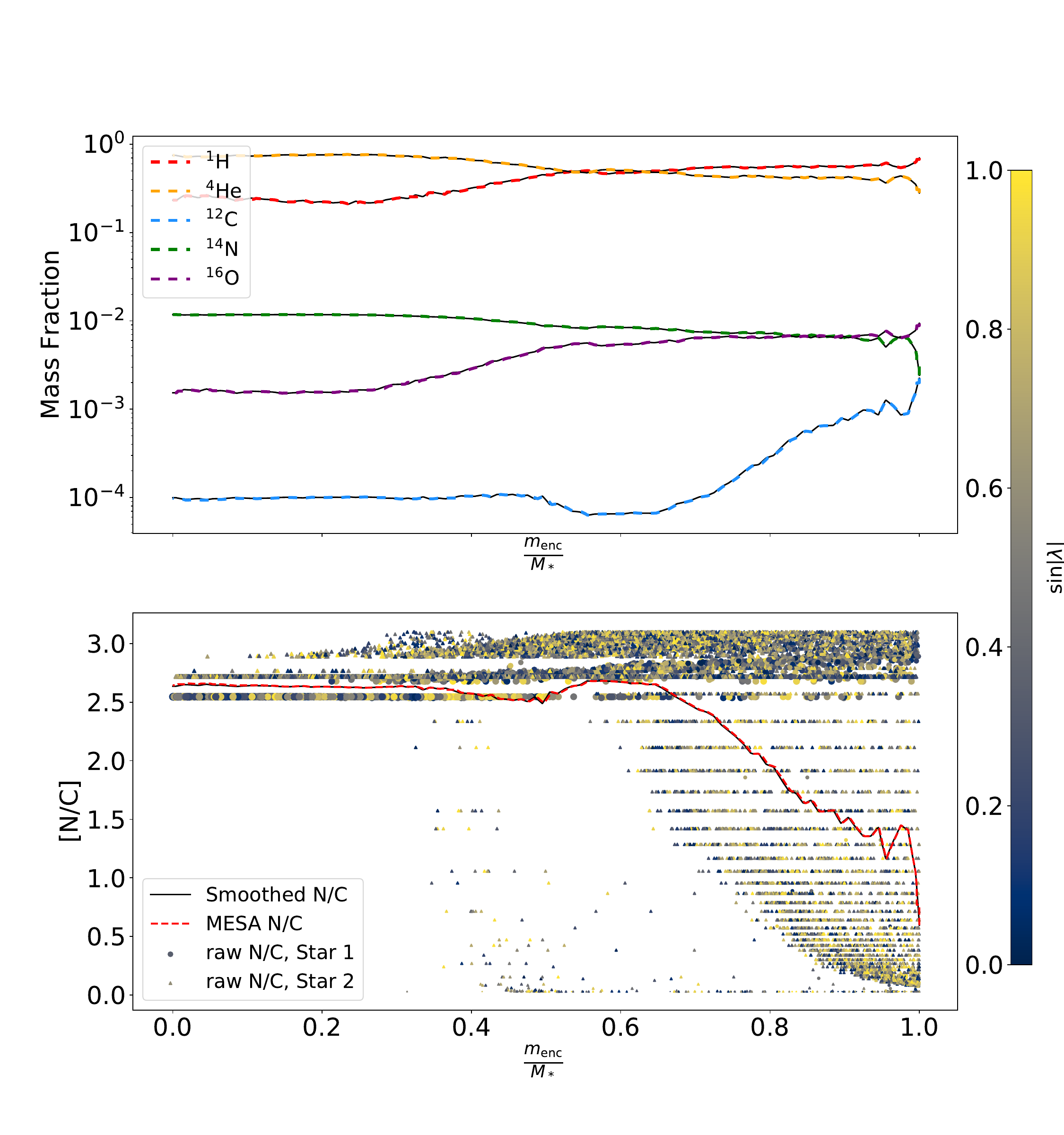}
        \caption{
        The composition profiles of the products of the $3\,M_{\odot}$ star for Case 32 (left) and the $3\,M_{\odot}$ star for the Case 55 (right). \textbf{(Top)} The solid curves show the results as given by SPH calculation being subsequently forced into HSE, while the dashed curves are for the {\tt MESA} model at the first output file from the {\tt MESA} relaxation. The profiles change very little during the contraction. \textbf{(Bottom)} The [N/C] value for several models of the product are shown. Each point represents the N/C quantity of every SPH particle for this product.  The color of each circular data point shows the sine of the magnitude of the latitude angle, $\sin\left|\lambda\right|$, of each particle in the product, and the area of each circle is proportional to the particle mass.
        The black, solid line shows the smoothed profile using the techniques described in Subsection \ref{subsec:sph_to_mesa}. The dashed, red line, showing high agreement with the smoothed SPH profile, is the N/C profile from the relaxed star in {\tt MESA}. The quantization of the [N/C] values in the raw SPH data is due to the particles having been placed on a lattice when initiating the parent star.}
        \label{fig:mesa_relax_composition}
\end{figure}

The mapping of our three-dimensional SPH models into one-dimensional sphericalized models, used in our simplified TDE treatment, necessarily discards some information.  While we expect differential rotation in a rotating product would cause mixing in the azimuthal direction due to smearing, a purely kinematic effect,
mixing along the polar angle is not necessarily expected.  
Information about the anisotropy of the compositional distribution and the particle distribution both from rotation and the asymmetry of the collision can be seen in Figure \ref{fig:mesa_relax_composition}, highlighting the simplifications to the stellar model described earlier.

\begin{figure}
    \centering
    \includegraphics[width=\linewidth]{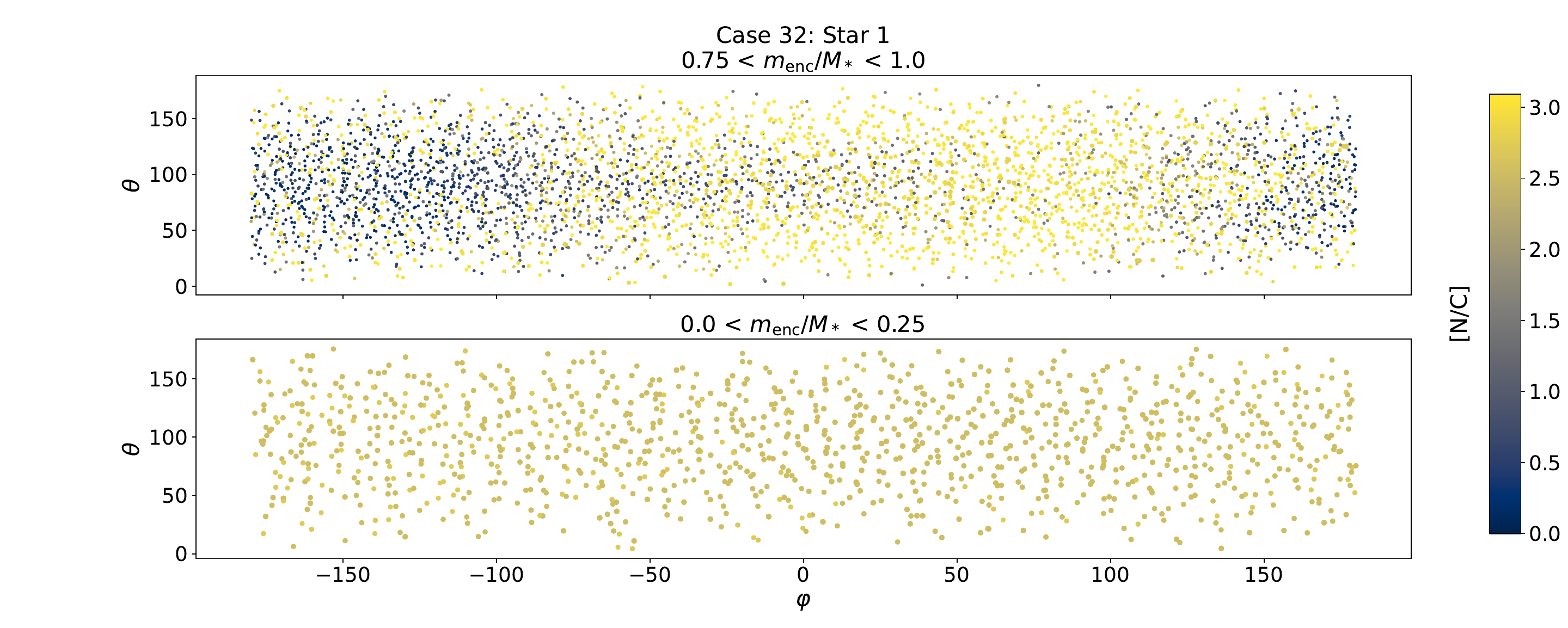}
    \caption{The angular spread of the composition of particles in the collision product of Star 1 from Case 32 at various radial regions. Each particle's size varies based on the mass of the particle; more massive particles are plotted to be larger.}
    \label{fig:angular_spread}
\end{figure}
Figure \ref{fig:angular_spread} shows anisotropies in the stellar composition that are ignored in the construction of the 1D stellar profile used in Figure \ref{fig:tde_fallback}. It illustrates that high N/C material is not evenly distributed throughout the outer layer of the star. In the outer layers, where the enclosed mass fraction is greater than 75\%, there are distinct regions with high and low N/C material. Particles in the equatorial region of the star experience smearing due to the rotation of the star. Differential rotation would likely smear more fully in the $\varphi$ direction. However, SPH tends to make stars rotate rigidly more quickly than in reality.
The inner 25\% of the mass is more uniform in composition, so the rotation does not contribute to anisotropies in the stellar composition. 

\bibliography{sample631}{}
\bibliographystyle{aasjournal}

\end{document}